\documentclass[11pt]{article}
\usepackage[a4paper,margin=1in]{geometry}
\usepackage{booktabs,tabularx,graphicx,amsmath,amssymb,microtype,xcolor}
\usepackage[numbers,square,sort&compress]{natbib}
\usepackage{xurl}
\usepackage[hidelinks]{hyperref}
\usepackage{placeins}
\newcommand{\PaperTitle}{Probabilistic Relevance Decisions for Semantic Retrieval}
\newcommand{\DenseScore}{s_{\mathrm{dense}}}
\newcommand{\JevScore}{s_{\mathrm{J}}}
\newcommand{\OpenAIScore}{s_{\mathrm{O}}}

\newcommand{\FullBGERecall}{0.719889}
\newcommand{\FullBGENDCG}{0.641277}
\newcommand{\ControlledBGERecall}{0.930333}
\newcommand{\ControlledBGENDCG}{0.899245}
\newcommand{\ControlledJevRecall}{0.992667}
\newcommand{\ControlledJevNDCG}{0.979298}
\newcommand{\HybridRecall}{0.826111}
\newcommand{\HybridNDCG}{0.780478}
\newcommand{\DatasetDocuments}{5,183}
\newcommand{\DatasetQueries}{300}
\newcommand{\DatasetPositives}{339}
\newcommand{\DenseDimension}{1,024}
\newcommand{\DenseMaxLength}{8,192}
\newcommand{\DenseBatch}{4}
\newcommand{\DenseThreads}{4}
\newcommand{\JevModel}{jev-1.13.0}
\newcommand{\DecisionVersion}{jev-relevance-v1}
\newcommand{\Seed}{42}
\newcommand{\ControlledCandidates}{100}
\newcommand{\CandidateRecall}{0.873667}
\newcommand{\ControlledRecallDelta}{0.062333}
\newcommand{\ControlledRecallLower}{0.035000}
\newcommand{\ControlledRecallUpper}{0.092000}
\newcommand{\ControlledNDCGDelta}{0.080054}
\newcommand{\ControlledNDCGLower}{0.055823}
\newcommand{\ControlledNDCGUpper}{0.105680}
\newcommand{\ControlledDecisions}{30,000}
\newcommand{\ControlledInput}{22,209,651}
\newcommand{\ControlledOutput}{690,000}
\newcommand{\ControlledCost}{0.932805342}
\newcommand{\ControlledQueryCost}{0.003109351}
\newcommand{\ControlledNewSpend}{0.871016832}
\newcommand{\HybridRecallDelta}{0.106222}
\newcommand{\HybridRecallLower}{0.070278}
\newcommand{\HybridRecallUpper}{0.142889}
\newcommand{\HybridNDCGDelta}{0.139201}
\newcommand{\HybridNDCGLower}{0.107902}
\newcommand{\HybridNDCGUpper}{0.170700}
\newcommand{\HybridDecisions}{15,000}
\newcommand{\HybridInput}{11,097,751}
\newcommand{\HybridOutput}{345,000}
\newcommand{\HybridCost}{0.466105542}
\newcommand{\HybridQueryCost}{0.001553685}
\newcommand{\HybridNewSpend}{0.447388746}
\newcommand{\FullBGELatencyMS}{55.641}
\newcommand{\HybridRerankLatencySeconds}{3.130931}
\newcommand{\BootstrapSamples}{10,000}

\newcommand{\ControlledOpenAIRecall}{0.960000}
\newcommand{\ControlledOpenAINDCG}{0.946291}
\newcommand{\ControlledOpenAIDecisions}{30,000}
\newcommand{\ControlledOpenAIInput}{17,016,283}
\newcommand{\ControlledOpenAIOutput}{0}
\newcommand{\ControlledOpenAICost}{1.701628300}
\newcommand{\ControlledOpenAIFreshCost}{1.589172300}
\newcommand{\ControlledOpenAIQueryCost}{0.005672094}
\newcommand{\ControlledOpenAIQueryMean}{8.038126}
\newcommand{\ControlledOpenAIQueryMedian}{7.997592}
\newcommand{\ControlledOpenAIQueryPninetyfive}{11.236097}
\newcommand{\ControlledOpenAITieRecallPrimary}{0.960000}
\newcommand{\ControlledOpenAITieRecallMean}{0.958333}
\newcommand{\ControlledOpenAITieRecallMin}{0.956667}
\newcommand{\ControlledOpenAITieRecallMax}{0.970000}
\newcommand{\ControlledOpenAITieRecallAffected}{12}
\newcommand{\ControlledOpenAITieNDCGPrimary}{0.946291}
\newcommand{\ControlledOpenAITieNDCGMean}{0.945004}
\newcommand{\ControlledOpenAITieNDCGMin}{0.942262}
\newcommand{\ControlledOpenAITieNDCGMax}{0.954624}
\newcommand{\ControlledOpenAITieNDCGAffected}{17}
\newcommand{\HybridOpenAIRecall}{0.802778}
\newcommand{\HybridOpenAINDCG}{0.754474}
\newcommand{\AbstractHybridOpenAINDCG}{0.7545}
\newcommand{\HybridOpenAIDecisions}{15,000}
\newcommand{\HybridOpenAIInput}{8,489,801}
\newcommand{\HybridOpenAIOutput}{0}
\newcommand{\HybridOpenAICost}{0.848980100}
\newcommand{\HybridOpenAIFreshCost}{0.814779100}
\newcommand{\HybridOpenAIQueryCost}{0.002829934}
\newcommand{\HybridOpenAIQueryMean}{4.875841}
\newcommand{\HybridOpenAIQueryMedian}{4.581425}
\newcommand{\HybridOpenAIQueryPninetyfive}{7.327947}
\newcommand{\HybridOpenAIComponentMean}{4.950946}
\newcommand{\HybridOpenAIComponentMedian}{4.664519}
\newcommand{\HybridOpenAIComponentPninetyfive}{7.389957}
\newcommand{\HybridOpenAITieRecallPrimary}{0.802778}
\newcommand{\HybridOpenAITieRecallMean}{0.798976}
\newcommand{\HybridOpenAITieRecallMin}{0.793778}
\newcommand{\HybridOpenAITieRecallMax}{0.805278}
\newcommand{\HybridOpenAITieRecallAffected}{9}
\newcommand{\HybridOpenAITieNDCGPrimary}{0.754474}
\newcommand{\HybridOpenAITieNDCGMean}{0.751919}
\newcommand{\HybridOpenAITieNDCGMin}{0.747658}
\newcommand{\HybridOpenAITieNDCGMax}{0.757911}
\newcommand{\HybridOpenAITieNDCGAffected}{43}
\newcommand{\JevControlledZero}{0.000000}
\newcommand{\JevControlledOne}{0.000000}
\newcommand{\JevControlledDistinct}{96}
\newcommand{\JevControlledBoundaryFive}{177}
\newcommand{\JevControlledBoundaryTen}{251}
\newcommand{\JevControlledMixedTen}{3}
\newcommand{\JevControlledPositiveMean}{0.784336}
\newcommand{\JevControlledNonpositiveMean}{0.020282}
\newcommand{\JevHybridZero}{0.000000}
\newcommand{\JevHybridOne}{0.000000}
\newcommand{\JevHybridDistinct}{99}
\newcommand{\JevHybridBoundaryFive}{86}
\newcommand{\JevHybridBoundaryTen}{151}
\newcommand{\JevHybridMixedTen}{18}
\newcommand{\JevHybridPositiveMean}{0.837987}
\newcommand{\JevHybridNonpositiveMean}{0.097372}
\newcommand{\OpenAIControlledZero}{0.974567}
\newcommand{\OpenAIControlledOne}{0.006200}
\newcommand{\OpenAIControlledDistinct}{94}
\newcommand{\OpenAIControlledBoundaryFive}{268}
\newcommand{\OpenAIControlledBoundaryTen}{289}
\newcommand{\OpenAIControlledMixedTen}{17}
\newcommand{\OpenAIControlledMixedBoundaryFive}{12}
\newcommand{\OpenAIControlledMixedBoundaryTen}{12}
\newcommand{\OpenAIControlledPositiveCount}{339}
\newcommand{\OpenAIControlledPositiveMean}{0.812566}
\newcommand{\OpenAIControlledNonpositiveMean}{0.002217}
\newcommand{\OpenAIHybridZero}{0.783533}
\newcommand{\OpenAIHybridOne}{0.020733}
\newcommand{\OpenAIHybridDistinct}{101}
\newcommand{\OpenAIHybridBoundaryFive}{145}
\newcommand{\OpenAIHybridBoundaryTen}{222}
\newcommand{\OpenAIHybridMixedTen}{43}
\newcommand{\OpenAIHybridMixedBoundaryFive}{9}
\newcommand{\OpenAIHybridMixedBoundaryTen}{4}
\newcommand{\OpenAIHybridPositiveCount}{298}
\newcommand{\OpenAIHybridPositiveMean}{0.869497}
\newcommand{\OpenAIHybridNonpositiveMean}{0.059695}
\newcommand{\ControlledOpenAIBGERecallDelta}{0.029667}
\newcommand{\ControlledOpenAIBGERecallLower}{-0.000333}
\newcommand{\ControlledOpenAIBGERecallUpper}{0.060667}
\newcommand{\ControlledOpenAIBGENDCGDelta}{0.047046}
\newcommand{\ControlledOpenAIBGENDCGLower}{0.019533}
\newcommand{\ControlledOpenAIBGENDCGUpper}{0.074930}
\newcommand{\ControlledOpenAIJevRecallDelta}{-0.032667}
\newcommand{\ControlledOpenAIJevRecallLower}{-0.054667}
\newcommand{\ControlledOpenAIJevRecallUpper}{-0.013333}
\newcommand{\ControlledOpenAIJevNDCGDelta}{-0.033008}
\newcommand{\ControlledOpenAIJevNDCGLower}{-0.054355}
\newcommand{\ControlledOpenAIJevNDCGUpper}{-0.013860}
\newcommand{\HybridOpenAIBGERecallDelta}{0.082889}
\newcommand{\HybridOpenAIBGERecallLower}{0.045111}
\newcommand{\HybridOpenAIBGERecallUpper}{0.120779}
\newcommand{\HybridOpenAIBGENDCGDelta}{0.113198}
\newcommand{\HybridOpenAIBGENDCGLower}{0.079198}
\newcommand{\HybridOpenAIBGENDCGUpper}{0.147932}
\newcommand{\HybridOpenAIJevRecallDelta}{-0.023333}
\newcommand{\HybridOpenAIJevRecallLower}{-0.047333}
\newcommand{\HybridOpenAIJevRecallUpper}{-0.000333}
\newcommand{\HybridOpenAIJevNDCGDelta}{-0.026003}
\newcommand{\HybridOpenAIJevNDCGLower}{-0.043124}
\newcommand{\HybridOpenAIJevNDCGUpper}{-0.008821}
\newcommand{\OpenAIModel}{gpt-6-luna}
\newcommand{\ExtensionFreshCost}{2.403951400}
\newcommand{\ExtensionFreshDecisions}{42,413}
\newcommand{\TieSeeds}{100}
\newcommand{\OpenAIInputRate}{0.100}
\newcommand{\OpenAIOutputRate}{0.000}
\newcommand{\JevInputRate}{0.042}
\newcommand{\JevOutputRate}{0.000}
\newcommand{\OriginalDenseControlledTotalSecondsMean}{0.061645}
\newcommand{\OriginalDenseControlledTotalSecondsMedian}{0.053524}
\newcommand{\OriginalDenseControlledTotalSecondsPninetyfive}{0.104555}
\newcommand{\OriginalJevControlledPerQueryLatencyMean}{6.008466}
\newcommand{\OriginalJevControlledPerQueryLatencyMedian}{6.230332}
\newcommand{\OriginalJevControlledPerQueryLatencyPninetyfive}{6.771913}
\newcommand{\OriginalDenseFullTotalSecondsMean}{0.075105}
\newcommand{\OriginalDenseFullTotalSecondsMedian}{0.055641}
\newcommand{\OriginalDenseFullTotalSecondsPninetyfive}{0.167162}
\newcommand{\OriginalHybridJevRerankingSecondsMean}{3.195131}
\newcommand{\OriginalHybridJevRerankingSecondsMedian}{3.130931}
\newcommand{\OriginalHybridJevRerankingSecondsPninetyfive}{3.489085}
\newcommand{\OriginalHybridComponentSumSecondsMean}{3.270236}
\newcommand{\OriginalHybridComponentSumSecondsMedian}{3.194208}
\newcommand{\OriginalHybridComponentSumSecondsPninetyfive}{3.626392}

\newcommand{\DepthFifty}{0.873667}
\newcommand{\DepthHundred}{0.897000}
\newcommand{\DepthAddedPairs}{8}
\newcommand{\DepthGainingQueries}{8}
\newcommand{\DepthCompleteFifty}{259}
\newcommand{\DepthMissingHundred}{33}
\newcommand{\DepthJevRecall}{0.831111}
\newcommand{\DepthJevTieRecallPrimary}{0.831111}
\newcommand{\DepthJevTieRecallMean}{0.831078}
\newcommand{\DepthJevTieRecallMin}{0.827778}
\newcommand{\DepthJevTieRecallMax}{0.834444}
\newcommand{\DepthJevTieRecallChanged}{2}
\newcommand{\DepthJevNDCG}{0.785189}
\newcommand{\DepthJevTieNDCGPrimary}{0.785189}
\newcommand{\DepthJevTieNDCGMean}{0.787368}
\newcommand{\DepthJevTieNDCGMin}{0.784475}
\newcommand{\DepthJevTieNDCGMax}{0.790340}
\newcommand{\DepthJevTieNDCGChanged}{19}
\newcommand{\DepthJevZero}{0.000000}
\newcommand{\DepthJevOne}{0.000000}
\newcommand{\DepthJevDistinct}{99}
\newcommand{\DepthJevTiedPairs}{0.218218}
\newcommand{\DepthJevCutoffFive}{74}
\newcommand{\DepthJevCutoffTen}{147}
\newcommand{\DepthJevMixedTen}{19}
\newcommand{\DepthJevImpactFive}{2}
\newcommand{\DepthJevImpactTen}{2}
\newcommand{\DepthJevCached}{15,323}
\newcommand{\DepthJevNew}{14,677}
\newcommand{\DepthJevCost}{0.933345924}
\newcommand{\DepthJevIncremental}{0.456901410}
\newcommand{\DepthJevQueryCost}{0.003111153}
\newcommand{\DepthJevCacheFraction}{0.510767}
\newcommand{\DepthJevRequestMean}{0.340930}
\newcommand{\DepthJevRequestMedian}{0.321688}
\newcommand{\DepthJevRequestPninetyfive}{0.447637}
\newcommand{\DepthJevScoringMean}{3.706196}
\newcommand{\DepthJevScoringMedian}{3.627568}
\newcommand{\DepthJevScoringPninetyfive}{4.207566}
\newcommand{\DepthJevCacheMean}{0.117414}
\newcommand{\DepthJevCacheMedian}{0.096720}
\newcommand{\DepthJevCachePninetyfive}{0.243263}
\newcommand{\DepthJevAssemblyMean}{0.001352}
\newcommand{\DepthJevAssemblyMedian}{0.001057}
\newcommand{\DepthJevAssemblyPninetyfive}{0.003356}
\newcommand{\DepthJevMixedMean}{3.951284}
\newcommand{\DepthJevMixedMedian}{3.852252}
\newcommand{\DepthJevMixedPninetyfive}{4.453214}
\newcommand{\DepthJevReplayMean}{0.111304}
\newcommand{\DepthJevReplayMedian}{0.073373}
\newcommand{\DepthJevReplayPninetyfive}{0.283693}
\newcommand{\DepthOpenAIRecall}{0.808778}
\newcommand{\DepthOpenAITieRecallPrimary}{0.808778}
\newcommand{\DepthOpenAITieRecallMean}{0.804446}
\newcommand{\DepthOpenAITieRecallMin}{0.798111}
\newcommand{\DepthOpenAITieRecallMax}{0.814944}
\newcommand{\DepthOpenAITieRecallChanged}{11}
\newcommand{\DepthOpenAINDCG}{0.753010}
\newcommand{\DepthOpenAITieNDCGPrimary}{0.753010}
\newcommand{\DepthOpenAITieNDCGMean}{0.750469}
\newcommand{\DepthOpenAITieNDCGMin}{0.743759}
\newcommand{\DepthOpenAITieNDCGMax}{0.757893}
\newcommand{\DepthOpenAITieNDCGChanged}{45}
\newcommand{\DepthOpenAIZero}{0.839100}
\newcommand{\DepthOpenAIOne}{0.010967}
\newcommand{\DepthOpenAIDistinct}{101}
\newcommand{\DepthOpenAITiedPairs}{0.739036}
\newcommand{\DepthOpenAICutoffFive}{125}
\newcommand{\DepthOpenAICutoffTen}{203}
\newcommand{\DepthOpenAIMixedTen}{45}
\newcommand{\DepthOpenAIImpactFive}{11}
\newcommand{\DepthOpenAIImpactTen}{7}
\newcommand{\DepthOpenAICached}{15,323}
\newcommand{\DepthOpenAINew}{14,677}
\newcommand{\DepthOpenAICost}{1.700444800}
\newcommand{\DepthOpenAIIncremental}{0.832519500}
\newcommand{\DepthOpenAIQueryCost}{0.005668149}
\newcommand{\DepthOpenAIRequestMean}{0.215214}
\newcommand{\DepthOpenAIRequestMedian}{0.208791}
\newcommand{\DepthOpenAIRequestPninetyfive}{0.309954}
\newcommand{\DepthOpenAIScoringMean}{3.983614}
\newcommand{\DepthOpenAIScoringMedian}{3.844398}
\newcommand{\DepthOpenAIScoringPninetyfive}{4.848743}
\newcommand{\DepthOpenAICacheMean}{0.088624}
\newcommand{\DepthOpenAICacheMedian}{0.084079}
\newcommand{\DepthOpenAICachePninetyfive}{0.161715}
\newcommand{\DepthOpenAIAssemblyMean}{0.001798}
\newcommand{\DepthOpenAIAssemblyMedian}{0.001512}
\newcommand{\DepthOpenAIAssemblyPninetyfive}{0.003850}
\newcommand{\DepthOpenAIMixedMean}{4.218767}
\newcommand{\DepthOpenAIMixedMedian}{4.110020}
\newcommand{\DepthOpenAIMixedPninetyfive}{4.999379}
\newcommand{\DepthOpenAIReplayMean}{0.089437}
\newcommand{\DepthOpenAIReplayMedian}{0.049498}
\newcommand{\DepthOpenAIReplayPninetyfive}{0.264061}
\newcommand{\DepthContrastADelta}{0.005000}
\newcommand{\DepthContrastALower}{-0.013333}
\newcommand{\DepthContrastAUpper}{0.023333}
\newcommand{\DepthContrastAImproved}{5}
\newcommand{\DepthContrastADeclined}{3}
\newcommand{\DepthContrastATied}{292}
\newcommand{\DepthContrastBDelta}{0.004711}
\newcommand{\DepthContrastBLower}{-0.005309}
\newcommand{\DepthContrastBUpper}{0.017154}
\newcommand{\DepthContrastBImproved}{6}
\newcommand{\DepthContrastBDeclined}{16}
\newcommand{\DepthContrastBTied}{278}
\newcommand{\DepthContrastCDelta}{0.006000}
\newcommand{\DepthContrastCLower}{-0.007333}
\newcommand{\DepthContrastCUpper}{0.020333}
\newcommand{\DepthContrastCImproved}{4}
\newcommand{\DepthContrastCDeclined}{3}
\newcommand{\DepthContrastCTied}{293}
\newcommand{\DepthContrastDDelta}{-0.001465}
\newcommand{\DepthContrastDLower}{-0.012051}
\newcommand{\DepthContrastDUpper}{0.010859}
\newcommand{\DepthContrastDImproved}{7}
\newcommand{\DepthContrastDDeclined}{22}
\newcommand{\DepthContrastDTied}{271}
\newcommand{\DepthContrastEDelta}{0.111222}
\newcommand{\DepthContrastELower}{0.073164}
\newcommand{\DepthContrastEUpper}{0.150000}
\newcommand{\DepthContrastEImproved}{43}
\newcommand{\DepthContrastEDeclined}{5}
\newcommand{\DepthContrastETied}{252}
\newcommand{\DepthContrastFDelta}{0.143912}
\newcommand{\DepthContrastFLower}{0.110982}
\newcommand{\DepthContrastFUpper}{0.176689}
\newcommand{\DepthContrastFImproved}{102}
\newcommand{\DepthContrastFDeclined}{21}
\newcommand{\DepthContrastFTied}{177}
\newcommand{\DepthContrastGDelta}{0.088889}
\newcommand{\DepthContrastGLower}{0.047556}
\newcommand{\DepthContrastGUpper}{0.129724}
\newcommand{\DepthContrastGImproved}{40}
\newcommand{\DepthContrastGDeclined}{10}
\newcommand{\DepthContrastGTied}{250}
\newcommand{\DepthContrastHDelta}{0.111733}
\newcommand{\DepthContrastHLower}{0.076422}
\newcommand{\DepthContrastHUpper}{0.147007}
\newcommand{\DepthContrastHImproved}{99}
\newcommand{\DepthContrastHDeclined}{36}
\newcommand{\DepthContrastHTied}{165}
\newcommand{\DepthContrastIDelta}{-0.022333}
\newcommand{\DepthContrastILower}{-0.048000}
\newcommand{\DepthContrastIUpper}{0.002000}
\newcommand{\DepthContrastIImproved}{6}
\newcommand{\DepthContrastIDeclined}{13}
\newcommand{\DepthContrastITied}{281}
\newcommand{\DepthContrastJDelta}{-0.032179}
\newcommand{\DepthContrastJLower}{-0.051530}
\newcommand{\DepthContrastJUpper}{-0.012272}
\newcommand{\DepthContrastJImproved}{23}
\newcommand{\DepthContrastJDeclined}{53}
\newcommand{\DepthContrastJTied}{224}

\newcommand{\AbstractFullBGENDCG}{0.6413}
\newcommand{\AbstractHybridNDCG}{0.7805}

\hypersetup{pdftitle={Probabilistic Relevance Decisions for Semantic Retrieval}}
\title{\PaperTitle}
\author{
Bahad{\i}r Tatar\textsuperscript{1,*} and Cevat \c{S}ener\textsuperscript{1,2} \\[0.3em]
\small \textsuperscript{1}Recruspace LLC \\[0.3em]
\small \textsuperscript{2}Middle East Technical University \\[0.3em]
\small \textsuperscript{*}Corresponding author: \href{mailto:bahadirtatar@recruspace.com}{bahadirtatar@recruspace.com}
}
\date{}
\begin{document}
\maketitle
\begin{abstract}
Dense similarity supplies reusable corpus representations, while probabilistic relevance decisions offer query-dependent document scoring. We compare dense-only BGE-M3 with TypeSafe Jev Noul and OpenAI Decisions Predicate scores on the official BEIR SciFact test benchmark. The study evaluates \DatasetQueries{} official test queries using identical positive-enriched \ControlledCandidates{}-document pools and full-corpus dense retrieval followed by reranking at candidate depths of 50 and 100. Unlike the controlled pools, neither dense prefix receives positive injection. Jev has the highest observed Recall@5 and nDCG@10 under the evaluated conditions. At depth 50, full-corpus nDCG@10 is \AbstractFullBGENDCG{} for BGE-M3, \AbstractHybridOpenAINDCG{} with OpenAI and \AbstractHybridNDCG{} with Jev. Extending depth from 50 to 100 increases known-positive candidate recall but produces small, uncertain final-effectiveness changes. All four paired-bootstrap depth intervals include zero. Unique-condition API scoring cost approximately doubles, while compatible cached decisions reduce incremental expenditure. OpenAI returns many exact-zero scores, and secondary exact-tie randomization reveals ranking sensitivity without replacing deterministic primary results. Top-100 latency mixes cached and new execution. Historical hybrid latency is a saved-component sum under different execution environments and timer boundaries. These exploratory, unadjusted analyses support relevance decisions as a complement to dense retrieval within this protocol, without establishing general superiority, probability calibration or equivalence between depths.

\end{abstract}
\section{Introduction}\label{sec:introduction}
Dense semantic retrieval compares separately encoded queries and documents, making a large corpus reusable across queries \cite{karpukhin2020dpr}. Sentence embeddings also support cosine-based semantic similarity \cite{reimers2019sbert}. Similarity is a representation-based ranking signal. It need not directly express whether a document contains evidence relevant to a particular scientific claim. Retrieve-then-rerank methods allocate additional query-dependent computation to a shortlist \cite{nogueira2019bert}, while retaining the first stage's candidate-recall constraint.

Sequence-to-sequence ranking already uses relevance-label token scores to order documents \cite{nogueira2020seq2seq}. Decision APIs provide another externally hosted scoring interface. In our experiments, Jev's Noul interface and OpenAI Decisions' Predicate interface each return a raw probability for an independent query--document decision. We ask each whether a document contains information relevant to assessing a claim, including supporting or contradicting evidence. The comparison concerns interfaces and observed outputs, rather than identical architectures or a verified common calibration scale. There is no answer-generation stage.

We evaluate the official BEIR SciFact test benchmark \cite{thakur2021beir,wadden2020scifact} with a dense-only BGE-M3 baseline \cite{chen2024bgem3}. Two complementary protocols distinguish ranking from discovery. We evaluate identical positive-enriched \ControlledCandidates{}-document pools for \DatasetQueries{} queries and rerank unchanged Top-50 and Top-100 prefixes retrieved from the full \DatasetDocuments{}-document corpus. Additional provider contrasts and candidate-depth analyses are exploratory and use unadjusted intervals.

Candidate depth controls known-evidence availability, the number of potentially expensive decisions, and opportunities for ranking errors or score ties. An exploratory, non-preregistered depth comparison extends the saved dense prefix from 50 to 100 without positive injection.

Our questions are: \textbf{RQ1}, how do dense similarity, Jev Noul and OpenAI Predicate rank identical bounded candidate pools? \textbf{RQ2}, how effectively do the two APIs rerank identical candidates retrieved by full-corpus BGE-M3? \textbf{RQ3}, how do score saturation, exact ties, usage-derived cost and operational latency affect interpretation and practical applicability? \textbf{RQ4}, how does increasing first-stage dense candidate depth from 50 to 100 affect reranking effectiveness, candidate recall, score behavior and operational cost?

Jev yields the highest observed metrics in the controlled and evaluated full-corpus conditions. In the full-corpus pipeline, nDCG@10 increases from \FullBGENDCG{} for dense retrieval to \HybridOpenAINDCG{} with OpenAI and \HybridNDCG{} with Jev at Top-50. Recall@5 changes from \FullBGERecall{} to \HybridOpenAIRecall{} and \HybridRecall{}, respectively. Controlled OpenAI Recall@5 uncertainty includes zero relative to BGE despite a positive point estimate. Its many exact-zero scores motivate a secondary tie analysis, without replacing deterministic primary rankings.

Our contribution is a shared-protocol empirical comparison, accompanied by query-level paired analysis, preserved raw scores, exact-tie sensitivity and traceable usage/timing accounting. We do not introduce two-stage retrieval, claim priority for decision-model comparison, or compare against a conventional cross-encoder. Hybrid Jev API cost is \HybridQueryCost{} USD/query and median reranking latency \HybridRerankLatencySeconds{} seconds, alongside a saved dense median of \FullBGELatencyMS{} milliseconds. These illustrate additional expenditure and latency under differing local/remote execution conditions, not portable model-speed estimates. One dataset, incomplete judgments and multiple unadjusted exploratory intervals constrain the conclusions.

\section{Related Work}\label{sec:related}
\subsection{Dense retrieval and semantic representations}
Separately encoded representations permit similarity computation without joint inference for every query--document pair. Sentence-BERT trains sentence embeddings that can be compared by cosine similarity \cite{reimers2019sbert}. Dense Passage Retrieval learns separate question and passage encoders for retrieval \cite{karpukhin2020dpr}. These related designs illustrate reusable representations, but semantic-similarity training and retrieval supervision are distinct objectives.

A single-vector bottleneck does not imply uniformly poor generalization. Large dual encoders improve out-of-domain retrieval through greater model capacity and multi-stage training while retaining fixed-size dot-product scoring \cite{ni2022gtr}. This evidence motivates evaluating a concrete retriever and training configuration rather than attributing all observed ranking errors to vector similarity. BEIR exposes variation across retrieval tasks and domains \cite{thakur2021beir}. One dataset cannot establish universal model superiority. SciFact identifies scientific evidence supporting or refuting claims \cite{wadden2020scifact}. We use its BEIR retrieval judgments rather than generating veracity decisions or rationales.

M3-Embedding supports dense, sparse and multi-vector representations \cite{chen2024bgem3}. Our BGE-M3 baseline selects only its normalized dense representation. The study therefore evaluates a specific dense retrieval configuration rather than every capability of the embedding model. Sentence-BERT and the large dual encoders provide context, not additional experimental baselines.

\subsection{Query--document interaction and neural reranking}
Query-dependent ranking encompasses several computational arrangements. BERT passage reranking jointly represents a query and candidate passage to estimate relevance \cite{nogueira2019bert}. ColBERT instead performs late interaction between independently encoded token representations, preserving offline document encoding while retaining token-level matching \cite{khattab2020colbert}. ColBERTv2 combines residual compression and denoised supervision to address the quality and storage costs of this multi-vector approach \cite{santhanam2022colbertv2}. Joint encoding, late interaction and single-vector similarity allocate storage and query-time computation differently. They are not interchangeable algorithms.

Sequence-to-sequence document ranking trains a model to produce relevance-label tokens and normalizes the corresponding true/false logits to obtain a score for each candidate \cite{nogueira2020seq2seq}. Candidates are then ordered by those scores. This is an established connection between language-model outputs and independent relevance scoring, but it does not establish that target-token probabilities equal Jev Noul or OpenAI Predicate scores, share their calibration, or reveal their proprietary implementations.

RankT5 directly outputs numerical query--document ranking scores and explores pairwise and listwise ranking losses \cite{zhuang2023rankt5}. The distinction concerns the training objective as well as the inference output. An independently computed score can be learned using relations among candidates. Our API experiment observes independent decisions and their ranking effects. It does not identify either provider's training loss.

Language-model reranking also includes generating a permutation of a group of candidate passages. Sun et al. investigate this instructional ranking approach with ChatGPT and GPT-4 \cite{sun2023rankgpt}. A generated group ordering differs from sorting independently returned scalar relevance decisions. Their models, inputs and protocols are not experimentally interchangeable with the interfaces evaluated here. Language-model-based reranking thus predates this study. No priority claim is made.

A bounded first stage permits richer scoring without evaluating every corpus document, but constrains the evidence available to the reranker. Our BGE-M3 followed by Jev or OpenAI Decisions uses this established retrieve-then-rerank organization. Cross-encoders, late interaction, sequence-to-sequence ranking, RankT5 and generative LLM reranking provide context. None is an experimental comparator here, and we make no performance claim about either decision API relative to these methods.

\subsection{Decision-oriented evaluation and comparative positioning}
A scholarly preprint evaluates Jev on classification, inference and structured-decision benchmarks \cite{deusser2026jev}. Another studies decision-oriented recommendation reranking using controlled candidates derived from SASRec \cite{lyu2026decision}. The latter uses a distribution over candidate choices, whereas our Jev and OpenAI conditions independently score a binary relevance decision for each document. Both works are preprints, without verified peer review. They provide evaluation context rather than proof of calibration or retrieval performance for the scores studied here.

We evaluate dense vector similarity and two probabilistic decision-scoring systems under shared scientific retrieval conditions. Jev Noul and OpenAI Predicate receive semantically equivalent relevance questions and identical candidate sets, with fixed provider-specific inputs. This makes their observed ranking behavior comparable within the protocol without establishing identical proprietary architectures, training procedures or probability scales. Controlled pools isolate ordering when every known positive is available. Dense Top-50 and Top-100 prefixes test reranking under a first-stage coverage constraint. Returned score saturation, exact ties, paired effectiveness, usage-derived cost and separately defined operational timers are evaluated for both providers. The contribution is this empirical comparison, not a new reranking architecture or a claim of priority for decision-based relevance scoring. Calibration, cross-domain robustness and conventional cross-encoder comparisons remain outside the evaluated scope.

\section{Methodology}\label{sec:methodology}
\subsection{Study design}
We investigate how probabilistic relevance decisions compare with dense vector similarity for semantic retrieval, and whether they improve dense retrieval when used as a reranking stage. The evaluation concerns retrieval and ranking of documents for scientific claims. It contains no answer-generation stage. We evaluate BGE-M3 and two independent relevance scorers, Jev and OpenAI Decisions, under shared SciFact protocols. We distinguish ranking within a controlled candidate pool from discovery of candidates in the full corpus.

\subsection{Dataset and relevance judgments}
We use the official BEIR SciFact test benchmark \cite{thakur2021beir,wadden2020scifact}, with \DatasetDocuments{} documents, \DatasetQueries{} test queries and \DatasetPositives{} positive query--document relevance judgments (qrels). Query text is the scientific claim. Document input combines its title, a newline and its text. Raw corpus, query and qrel files were preserved unchanged, without alteration. Only the official test queries enter the evaluation.

We refer to pairs without a positive qrel as \emph{nonpositive}. This term describes the benchmark annotation status and does not establish that a document is irrelevant because many pairs are unjudged. Recall and nDCG are evaluated against the official qrels, with absent judgments contributing no positive gain under that reference.

\subsection{Dense retrieval baseline}
The dense baseline uses \texttt{BAAI/bge-m3} \cite{chen2024bgem3}. We use only the dense representation. CLS pooling followed by L2 normalization produces \DenseDimension{}-dimensional embeddings. Sparse representations and ColBERT/multi-vector functionality are excluded. For normalized embeddings $e_q$ and $e_d$, the score is
\begin{equation}
\DenseScore(q,d)=\langle e_q,e_d\rangle,
\end{equation}
which is cosine similarity for unit-length vectors. Claims receive no added query instruction. Documents receive no chunking or query-dependent expansion. The configured maximum input length is \DenseMaxLength{} tokens.

Inference uses float16 on Apple MPS, document batches of \DenseBatch{}, and \DenseThreads{} CPU threads. Encoded vectors are stored as float32. Corpus embeddings are cached, and CPU NumPy computes dot products and rankings using the stored matrix. Queries are encoded individually. All methods break exact score ties by lexicographic document ID after descending raw score. This is a deterministic rule on the string identifiers, not numeric identifier order. Bounded inference workers release persistent MPS allocations without changing model precision, text construction, maximum length or scoring.

\subsection{Probabilistic relevance decisions}
The Jev condition uses the pinned model \texttt{\JevModel} and the frozen specification \expandafter\nolinkurl\expandafter{\DecisionVersion}. Each claim--document pair is evaluated independently, with the claim, document title and text supplied as state. The question is whether the candidate contains information relevant to assessing the scientific claim. Its positive criterion includes information supporting, contradicting, qualifying or otherwise directly bearing on the claim. It does not ask whether the claim is true or supported specifically.

The ranking signal is the raw returned Noul relevance probability,
\begin{equation}
\JevScore(q,d)=P(R=1\mid q,d),
\end{equation}
where $R$ denotes document relevance under the decision specification. This notation represents the API's returned score. Probabilistic calibration was not evaluated. Candidates are sorted by this score in descending order. No relevance threshold was selected, and no decision wording was tuned after observing benchmark results. The relevance criterion remains fixed across both candidate protocols and depths. No claim about model internals is required by this evaluation.

\subsection{OpenAI Predicate scoring}
The OpenAI condition uses \texttt{\OpenAIModel} at \texttt{POST /v1/decisions}, with one named \texttt{predicate} question per claim--document request. A deterministic text input labels the claim, document title and document text, separated by newlines. The positive and negative instructions retain the Jev relevance criterion. This creates semantic task equivalence, not evidence of identical architectures, capabilities, output resolution or probability calibration.

We rank by the raw \texttt{probability} in the named predicate answer, $\OpenAIScore(q,d)$, independently of other candidates. We neither ask for explanations nor generate answers to claims. Requested and returned model identifiers, raw responses, request identities, serializer version and usage are preserved. A refusal or missing/invalid probability is a recorded failure, not an assigned zero. The final two conditions are complete. The wording was frozen after connectivity verification and before pilot/full evaluation. No quality-driven test-set prompt tuning was performed.

\subsection{Score diagnostics and exact-tie sensitivity}
Saved scores are summarized separately for each provider and candidate condition, including counts, means, medians, dispersion, exact-zero/one fractions and distinct values. Qrel-positive and nonpositive groups are descriptive benchmark strata. We distinguish ties anywhere in a pool, ties crossing the top-five or top-ten boundary, mixed-label tied groups intersecting the top ten, and changes in evaluated metrics. These definitions prevent a large nonpositive tied tail from being mistaken for an effectiveness loss.

The OpenAI secondary sensitivity analysis uses \TieSeeds{} predeclared seeds, 0--99. For each seed, a Python \texttt{random.Random(seed)} stream is reset for each query. Lexicographically sorted IDs are shuffled only within exactly equal raw-probability groups. Unequal scores retain their ordering. Official-qrel metrics are then averaged over the same test queries. Reported minima and maxima cover these seeded realizations, not every possible permutation. Primary rankings remain the original document-ID policy. No favorable seed is selected.

\subsection{Controlled candidate-pool evaluation}
For each of the \DatasetQueries{} queries, we construct one \ControlledCandidates{}-document pool containing every official positive plus sampled nonpositive corpus documents. Sampling uses seed \Seed{} and deterministic per-query streams. The candidate pools are fixed across all three scorers. Dense ranking did not select the sampled nonpositive candidates. Pilot candidate pools are retained in the controlled benchmark.

BGE-M3, Jev and OpenAI receive exactly the same frozen candidates. This condition isolates ranking quality given a bounded candidate set, with positives guaranteed to be present. It does not evaluate full-corpus Jev or OpenAI candidate discovery. The guarantee of positive inclusion distinguishes it from the full-corpus retrieval condition below.

\subsection{Full-corpus retrieval and decision-based reranking}
The full dense condition ranks all \DatasetDocuments{} SciFact documents for each of the \DatasetQueries{} official test queries. The saved BGE-M3 ranking defines candidate prefixes at depths 50 and 100, with the Top-50 list as an exact ordered prefix of Top-100. Within each depth, Jev and OpenAI receive the same candidate identities, score each claim--document pair independently and rerank by raw returned relevance probability. Top-50 remains the original condition, and the Top-100 extension is exploratory and non-preregistered. Compatible historical decisions were reused for the extension, as detailed in Section~\ref{sec:depthmethod}. No positives are injected into either dense prefix. Neither reranker can recover a document absent from its candidate set. Dense Recall@50 and Recall@100 therefore describe the respective candidate-stage recall ceilings. These retrieved prefixes differ from Controlled-100 pools, which guarantee inclusion of every known positive.

\subsection{Metrics and paired statistical analysis}
For a query with positive reference set $G_q$ and top-$k$ ranking $L_q^k$, recall is $|G_q\cap L_q^k|/|G_q|$. Primary outcomes are Recall@5 and nDCG@10. Cumulative-gain evaluation credits relevance at ranked positions and normalization against an ideal ordering \cite{jarvelin2002gain}. Our specific gain and discount definition follows. We compute $\mathrm{DCG}@k=\sum_{i=1}^k r_i/\log_2(i+1)$ with official graded qrel gain $r_i$, and normalize by the ideal DCG for the query. Query-level values are retained. Aggregate metrics are arithmetic means with equal query weight. Saved rankings and per-query and aggregate values were independently checked against \texttt{ir\_measures} using the official qrels.

In information retrieval (IR), statistical comparisons commonly use matched query-level outcomes and align inference with the reported statistic \cite{smucker2007statistics}. That methodological context does not validate our exact percentile intervals or multiplicity policy. Inference uses \DatasetQueries{} paired test queries and \BootstrapSamples{} paired bootstrap resamples, seed \Seed{}, and percentile 95\% confidence intervals. Each resample draws query indices with replacement and averages treatment-minus-baseline differences. Principal contrasts compare Jev against BGE-M3 in the controlled and full-corpus reranking conditions. Additional contrasts compare controlled OpenAI against BGE and Jev, and hybrid OpenAI against dense and Jev, each on both primary metrics. The additional comparisons are exploratory, and the intervals are unadjusted for multiplicity. A near-zero endpoint is reported without converting the analysis into a confirmatory general superiority claim. The pilot is not subjected to an additional significance test. We report observed differences and intervals rather than infer improvement solely from aggregate scores.

\subsection{Cost and operational latency}
Decision-API monetary cost is derived from API-returned input/output usage and a preserved, versioned pricing configuration. Condition-level successful-decision costs include reused decisions once, and fresh-call spend is logged separately. Shared decisions mean costs cannot simply be summed across conditions as fresh expenditure. Local BGE monetary cost was not measured. Usage for a failed HTTP response was not returned and remains unknown.

Latency is observed operational latency. BGE runs locally on macOS ARM/MPS with 8\,GB memory and float16 inference, whereas Jev and OpenAI use remote hosted APIs. Provider routes, retries and cache state differ. These environments do not support claims about hardware-controlled intrinsic model speed. Dense query encoding and ranking are timed separately from offline corpus construction and model loading. Each query worker has an excluded warm-up. Jev records successful request latency separately from complete query wall time, including cache lookup and persistence.

Controlled execution reused pilot decisions and required repair of one failed pair. Successful queries retain their original wall times. The repaired query sums active original and repair durations, excluding the gap between attempts. These timings describe the preserved cache/resume state, not uninterrupted fresh execution. Hybrid total latency is a component sum of saved dense query timing and measured reranking wall time, not a newly measured combined replay.

Candidate definitions, rankings, query-level metrics, raw Jev responses, usage, errors and manifests were preserved. OpenAI query ranking time includes candidate selection, cache lookup, scoring, response persistence, sorting and ranking persistence, but excludes metric evaluation. Scoring/persistence time and individual request/evaluation durations are separately retained. Earlier pilot timing excludes some cache and sorting work and is not relabeled as equivalent end-to-end time. Cache-only replay timing is not a fresh-inference measurement. Standard-rate estimates use the recorded execution-time pricing configurations, namely \JevInputRate{} USD per million input tokens for Jev and \OpenAIInputRate{} for OpenAI, with respective output rates \JevOutputRate{} and \OpenAIOutputRate{}. 

\subsection{Candidate-Depth Sensitivity Evaluation}\label{sec:depthmethod}
The exploratory, non-preregistered candidate-depth analysis reuses saved full-corpus BGE-M3 rankings to compare reranking depths 50 and 100. Each Top-100 candidate list is exactly the first 100 saved dense-ranked documents, with the Top-50 list as an ordered prefix. There is no resampling or positive injection. \textbf{Controlled-100} instead guarantees every known positive in deterministic sampled pools. It is a separate ranking protocol, not dense Top-100 retrieval.

Both providers retain their frozen models, relevance specifications and serializers. Compatible historical decisions were reused, and only missing query--document identities were scored. Raw scores are ranked descending with lexicographic document-ID ties. Neither historical scores nor ranking policies changed. Candidate Recall@50 and Recall@100 measure macro known-positive coverage before reranking, whereas Recall@5 and nDCG@10 measure final ordering. The depth analysis retains all \DatasetQueries{} official queries, \BootstrapSamples{} paired bootstrap resamples, seed \Seed{} and unadjusted 95\% percentile intervals. Secondary exact-tie sensitivity uses seeds 0--99, resetting \texttt{random.Random(seed)} per query and shuffling only equal-score groups after sorting their IDs. These are exploratory analyses, not a confirmatory family or an equivalence test.

Top-100 operational timers distinguish successful new HTTP requests, incremental scoring, cache lookup, ranking assembly, mixed cached/new query wall time and cache-only replay. Candidate loading, preflight audit and dense inference are excluded from the mixed query clock, which includes metric verification and serialization. Historical Top-50 timing boundaries differ. Unique-condition cost counts successful usage once per candidate identity, and incremental expenditure counts only newly successful decisions. 

\section{Results}\label{sec:results}
\subsection{Controlled candidate ranking}
Table~\ref{tab:controlled} evaluates the three scorers on identical positive-enriched pools. Jev has the highest observed Recall@5 and nDCG@10. OpenAI exceeds the dense baseline in both point estimates and remains below Jev. This evaluates bounded ordering rather than full-corpus Jev retrieval or OpenAI document discovery. The known-positive guarantee makes these scores unsuitable for direct comparison with the full-corpus condition.
\begin{table}[tb]\centering\small
\caption{Ranking identical positive-enriched 100-document pools for all 300 official test queries. This is bounded ranking, rather than independent full-corpus discovery.}\label{tab:controlled}
\begin{tabular}{@{}lrr@{}}\toprule
Method & Recall@5 & nDCG@10\\\midrule
BGE-M3 & \ControlledBGERecall{} & \ControlledBGENDCG{}\\
Jev & \ControlledJevRecall{} & \ControlledJevNDCG{}\\
OpenAI Decisions & \ControlledOpenAIRecall{} & \ControlledOpenAINDCG{}\\
\bottomrule\end{tabular}\end{table}

\subsection{Full-corpus dense retrieval and decision-based reranking}
Table~\ref{tab:hybrid} reports the practical full-corpus pipeline, including the Top-100 conditions discussed in Section~\ref{sec:depthresults}. The following contrasts use Top-50. Both APIs improve the ordering of the same saved dense top-50 lists, and Jev again has the highest observed effectiveness. Dense Recall@50 is \CandidateRecall{}, the shared candidate-stage ceiling. Neither reranker can recover a relevant document outside these lists, and no positives were injected. Full-corpus independent API scoring was not executed. The historical Jev brute-force cost projection remains labeled \texttt{PROJECTED / NOT EXECUTED}. It supplies no measured effectiveness or latency result.
\begin{table}[tbp]\centering\footnotesize
\caption{Full-corpus dense retrieval and exact dense-prefix reranking, without positive injection. Coverage is macro known-positive candidate recall; the full dense baseline ranks the entire corpus once. Controlled-100 pools are a separate protocol.}\label{tab:hybrid}
\begin{tabular}{@{}lrrrr@{}}\toprule
Method & Depth & Recall@5 & nDCG@10 & Candidate recall\\\midrule
BGE-M3 full corpus & \DatasetDocuments{} & \FullBGERecall{} & \FullBGENDCG{} & ---\\
BGE-M3 $\to$ Jev & 50 & \HybridRecall{} & \HybridNDCG{} & \DepthFifty{}\\
BGE-M3 $\to$ Jev & 100 & \DepthJevRecall{} & \DepthJevNDCG{} & \DepthHundred{}\\
BGE-M3 $\to$ OpenAI & 50 & \HybridOpenAIRecall{} & \HybridOpenAINDCG{} & \DepthFifty{}\\
BGE-M3 $\to$ OpenAI & 100 & \DepthOpenAIRecall{} & \DepthOpenAINDCG{} & \DepthHundred{}\\
\bottomrule\end{tabular}\end{table}

\subsection{Paired statistical comparisons}
For controlled pools and dense Top-50 reranking, Table~\ref{tab:bootstrap} reports Jev--BGE and exploratory OpenAI contrasts. Controlled OpenAI-minus-BGE nDCG@10 has a positive interval, whereas its Recall@5 interval, $[\ControlledOpenAIBGERecallLower{},\ControlledOpenAIBGERecallUpper{}]$, includes zero. Thus the positive Recall point estimate alone does not establish a clearly positive paired difference. Both hybrid OpenAI-minus-dense intervals are positive. OpenAI-minus-Jev intervals are negative in each condition, with hybrid Recall@5's upper endpoint \HybridOpenAIJevRecallUpper{} close to zero.

These are query-sample uncertainty estimates for the evaluated protocol. The additional contrasts are exploratory and the intervals are not multiplicity-adjusted. They do not establish general superiority across models, datasets or deployments. The controlled Recall uncertainty remains part of the interpretation.
\begin{table}[tb]\centering\footnotesize
\caption{Paired differences and unadjusted 95\% percentile intervals from 10,000 query-level resamples with seed 42. Principal and additional exploratory comparisons remain distinct.}\label{tab:bootstrap}
\begin{tabular}{@{}llrr@{}}\toprule
Comparison & Metric & $\Delta$ & 95\% interval\\\midrule
Principal: controlled Jev $-$ BGE & Recall@5 & \ControlledRecallDelta{} & $[\ControlledRecallLower{}, \ControlledRecallUpper{}]$\\
Principal: controlled Jev $-$ BGE & nDCG@10 & \ControlledNDCGDelta{} & $[\ControlledNDCGLower{}, \ControlledNDCGUpper{}]$\\
Principal: hybrid Jev $-$ BGE & Recall@5 & \HybridRecallDelta{} & $[\HybridRecallLower{}, \HybridRecallUpper{}]$\\
Principal: hybrid Jev $-$ BGE & nDCG@10 & \HybridNDCGDelta{} & $[\HybridNDCGLower{}, \HybridNDCGUpper{}]$\\
Exploratory: controlled OpenAI $-$ BGE & Recall@5 & \ControlledOpenAIBGERecallDelta{} & $[\ControlledOpenAIBGERecallLower{}, \ControlledOpenAIBGERecallUpper{}]$\\
Exploratory: controlled OpenAI $-$ BGE & nDCG@10 & \ControlledOpenAIBGENDCGDelta{} & $[\ControlledOpenAIBGENDCGLower{}, \ControlledOpenAIBGENDCGUpper{}]$\\
Exploratory: controlled OpenAI $-$ Jev & Recall@5 & \ControlledOpenAIJevRecallDelta{} & $[\ControlledOpenAIJevRecallLower{}, \ControlledOpenAIJevRecallUpper{}]$\\
Exploratory: controlled OpenAI $-$ Jev & nDCG@10 & \ControlledOpenAIJevNDCGDelta{} & $[\ControlledOpenAIJevNDCGLower{}, \ControlledOpenAIJevNDCGUpper{}]$\\
Exploratory: hybrid OpenAI $-$ BGE & Recall@5 & \HybridOpenAIBGERecallDelta{} & $[\HybridOpenAIBGERecallLower{}, \HybridOpenAIBGERecallUpper{}]$\\
Exploratory: hybrid OpenAI $-$ BGE & nDCG@10 & \HybridOpenAIBGENDCGDelta{} & $[\HybridOpenAIBGENDCGLower{}, \HybridOpenAIBGENDCGUpper{}]$\\
Exploratory: hybrid OpenAI $-$ Jev & Recall@5 & \HybridOpenAIJevRecallDelta{} & $[\HybridOpenAIJevRecallLower{}, \HybridOpenAIJevRecallUpper{}]$\\
Exploratory: hybrid OpenAI $-$ Jev & nDCG@10 & \HybridOpenAIJevNDCGDelta{} & $[\HybridOpenAIJevNDCGLower{}, \HybridOpenAIJevNDCGUpper{}]$\\
\bottomrule\end{tabular}\end{table}

\begin{table}[tbp]\centering\footnotesize
\caption{Usage-derived API costs in USD (panel A) and recorded operational latency in seconds (panel B).}
\label{tab:costlatency}\label{tab:usage}\label{tab:latency}
\textbf{A. Successful-decision usage and unique condition cost}\par\medskip
\begin{tabular}{@{}lrrrrr@{}}\toprule
Condition & Decisions & Input tokens & Output tokens & USD total & USD/query\\\midrule
Jev controlled & \ControlledDecisions{} & \ControlledInput{} & \ControlledOutput{} & \ControlledCost{} & \ControlledQueryCost{}\\
OpenAI controlled & \ControlledOpenAIDecisions{} & \ControlledOpenAIInput{} & \ControlledOpenAIOutput{} & \ControlledOpenAICost{} & \ControlledOpenAIQueryCost{}\\
Jev hybrid & \HybridDecisions{} & \HybridInput{} & \HybridOutput{} & \HybridCost{} & \HybridQueryCost{}\\
OpenAI hybrid & \HybridOpenAIDecisions{} & \HybridOpenAIInput{} & \HybridOpenAIOutput{} & \HybridOpenAICost{} & \HybridOpenAIQueryCost{}\\
\bottomrule\end{tabular}
\par\bigskip\textbf{B. Recorded query / reranking / component-sum timers}\par\medskip
\begin{tabular}{@{}lrrr@{}}\toprule
Condition / timer & Mean & Median & p95\\\midrule
BGE controlled & \OriginalDenseControlledTotalSecondsMean{} & \OriginalDenseControlledTotalSecondsMedian{} & \OriginalDenseControlledTotalSecondsPninetyfive{}\\
Jev controlled & \OriginalJevControlledPerQueryLatencyMean{} & \OriginalJevControlledPerQueryLatencyMedian{} & \OriginalJevControlledPerQueryLatencyPninetyfive{}\\
BGE full corpus & \OriginalDenseFullTotalSecondsMean{} & \OriginalDenseFullTotalSecondsMedian{} & \OriginalDenseFullTotalSecondsPninetyfive{}\\
Jev hybrid reranking & \OriginalHybridJevRerankingSecondsMean{} & \OriginalHybridJevRerankingSecondsMedian{} & \OriginalHybridJevRerankingSecondsPninetyfive{}\\
Jev hybrid component sum & \OriginalHybridComponentSumSecondsMean{} & \OriginalHybridComponentSumSecondsMedian{} & \OriginalHybridComponentSumSecondsPninetyfive{}\\
OpenAI controlled & \ControlledOpenAIQueryMean{} & \ControlledOpenAIQueryMedian{} & \ControlledOpenAIQueryPninetyfive{}\\
OpenAI hybrid reranking & \HybridOpenAIQueryMean{} & \HybridOpenAIQueryMedian{} & \HybridOpenAIQueryPninetyfive{}\\
OpenAI hybrid component sum & \HybridOpenAIComponentMean{} & \HybridOpenAIComponentMedian{} & \HybridOpenAIComponentPninetyfive{}\\
\bottomrule\end{tabular}
\end{table}

\subsection{Relevance-score distributions}
\begin{figure}[tb]\centering
\resizebox{\linewidth}{!}{%
\setlength{\unitlength}{1pt}
\begin{picture}(500,330)
\linethickness{0.6pt}
\color{black}
\put(55.000,195.000){\line(1,0){182.000}}
\put(55.000,195.000){\line(0,1){100.000}}
\put(146.000,307.000){\makebox(0,0)[c]{\small Controlled-100; nonpositive}}
\color[HTML]{536878}
\linethickness{1pt}
\put(55.000,195.000){\line(0,1){88.346}}
\put(55.000,283.346){\line(1,0){9.100}}
\put(64.100,198.451){\line(0,1){84.895}}
\put(64.100,198.451){\line(1,0){9.100}}
\put(73.200,195.704){\line(0,1){2.747}}
\put(73.200,195.704){\line(1,0){9.100}}
\put(82.300,195.226){\line(0,1){0.478}}
\put(82.300,195.226){\line(1,0){9.100}}
\put(91.400,195.116){\line(0,1){0.110}}
\put(91.400,195.116){\line(1,0){9.100}}
\put(100.500,195.116){\line(0,1){0.019}}
\put(100.500,195.135){\line(1,0){9.100}}
\put(109.600,195.035){\line(0,1){0.101}}
\put(109.600,195.035){\line(1,0){9.100}}
\put(118.700,195.025){\line(0,1){0.009}}
\put(118.700,195.025){\line(1,0){9.100}}
\put(127.800,195.025){\line(0,1){0.016}}
\put(127.800,195.041){\line(1,0){9.100}}
\put(136.900,195.016){\line(0,1){0.025}}
\put(136.900,195.016){\line(1,0){9.100}}
\put(146.000,195.016){\line(0,1){0.013}}
\put(146.000,195.028){\line(1,0){9.100}}
\put(155.100,195.013){\line(0,1){0.016}}
\put(155.100,195.013){\line(1,0){9.100}}
\put(164.200,195.013){\line(0,1){0.019}}
\put(164.200,195.031){\line(1,0){9.100}}
\put(173.300,195.019){\line(0,1){0.013}}
\put(173.300,195.019){\line(1,0){9.100}}
\put(182.400,195.006){\line(0,1){0.013}}
\put(182.400,195.006){\line(1,0){9.100}}
\put(191.500,195.006){\line(0,1){0.003}}
\put(191.500,195.009){\line(1,0){9.100}}
\put(200.600,195.009){\line(0,1){0.013}}
\put(200.600,195.022){\line(1,0){9.100}}
\put(209.700,195.000){\line(0,1){0.022}}
\put(209.700,195.000){\line(1,0){9.100}}
\put(218.800,195.000){\line(0,1){0.003}}
\put(218.800,195.003){\line(1,0){9.100}}
\put(227.900,195.003){\line(0,1){0.000}}
\put(227.900,195.003){\line(1,0){9.100}}
\put(237.000,195.000){\line(0,1){0.003}}
\color[HTML]{BF5B35}
\linethickness{1pt}
\put(55.000,195.000){\line(0,1){92.593}}
\put(55.000,287.593){\line(1,0){9.100}}
\put(64.100,195.154){\line(0,1){92.439}}
\put(64.100,195.154){\line(1,0){9.100}}
\put(73.200,195.101){\line(0,1){0.053}}
\put(73.200,195.101){\line(1,0){9.100}}
\put(82.300,195.060){\line(0,1){0.041}}
\put(82.300,195.060){\line(1,0){9.100}}
\put(91.400,195.053){\line(0,1){0.006}}
\put(91.400,195.053){\line(1,0){9.100}}
\put(100.500,195.053){\line(0,1){0.003}}
\put(100.500,195.057){\line(1,0){9.100}}
\put(109.600,195.035){\line(0,1){0.022}}
\put(109.600,195.035){\line(1,0){9.100}}
\put(118.700,195.013){\line(0,1){0.022}}
\put(118.700,195.013){\line(1,0){9.100}}
\put(127.800,195.013){\line(0,1){0.000}}
\put(127.800,195.013){\line(1,0){9.100}}
\put(136.900,195.009){\line(0,1){0.003}}
\put(136.900,195.009){\line(1,0){9.100}}
\put(146.000,195.009){\line(0,1){0.013}}
\put(146.000,195.022){\line(1,0){9.100}}
\put(155.100,195.013){\line(0,1){0.009}}
\put(155.100,195.013){\line(1,0){9.100}}
\put(164.200,195.009){\line(0,1){0.003}}
\put(164.200,195.009){\line(1,0){9.100}}
\put(173.300,195.009){\line(0,1){0.006}}
\put(173.300,195.016){\line(1,0){9.100}}
\put(182.400,195.016){\line(0,1){0.003}}
\put(182.400,195.019){\line(1,0){9.100}}
\put(191.500,195.019){\line(0,1){0.000}}
\put(191.500,195.019){\line(1,0){9.100}}
\put(200.600,195.009){\line(0,1){0.009}}
\put(200.600,195.009){\line(1,0){9.100}}
\put(209.700,195.000){\line(0,1){0.009}}
\put(209.700,195.000){\line(1,0){9.100}}
\put(218.800,195.000){\line(0,1){0.006}}
\put(218.800,195.006){\line(1,0){9.100}}
\put(227.900,195.006){\line(0,1){0.025}}
\put(227.900,195.031){\line(1,0){9.100}}
\put(237.000,195.000){\line(0,1){0.031}}
\color{black}
\put(12.000,245.000){\rotatebox{90}{\small Count}}
\color{black}
\put(146.000,172.000){\makebox(0,0)[c]{\small Raw probability}}
\put(55.000,185.000){\makebox(0,0)[c]{\small 0}}
\put(146.000,185.000){\makebox(0,0)[c]{\small 0.5}}
\put(237.000,185.000){\makebox(0,0)[c]{\small 1}}
\put(50.000,195.000){\makebox(0,0)[r]{\small 0}}
\put(50.000,245.000){\makebox(0,0)[r]{\small 15907}}
\put(50.000,295.000){\makebox(0,0)[r]{\small 31815}}
\color{black}
\put(305.000,195.000){\line(1,0){182.000}}
\put(305.000,195.000){\line(0,1){100.000}}
\put(396.000,307.000){\makebox(0,0)[c]{\small Controlled-100; qrel positive}}
\color[HTML]{536878}
\linethickness{1pt}
\put(305.000,195.000){\line(0,1){0.441}}
\put(305.000,195.441){\line(1,0){9.100}}
\put(314.100,195.441){\line(0,1){3.086}}
\put(314.100,198.527){\line(1,0){9.100}}
\put(323.200,198.086){\line(0,1){0.441}}
\put(323.200,198.086){\line(1,0){9.100}}
\put(332.300,197.646){\line(0,1){0.441}}
\put(332.300,197.646){\line(1,0){9.100}}
\put(341.400,197.646){\line(0,1){0.882}}
\put(341.400,198.527){\line(1,0){9.100}}
\put(350.500,198.527){\line(0,1){0.882}}
\put(350.500,199.409){\line(1,0){9.100}}
\put(359.600,198.086){\line(0,1){1.323}}
\put(359.600,198.086){\line(1,0){9.100}}
\put(368.700,198.086){\line(0,1){0.882}}
\put(368.700,198.968){\line(1,0){9.100}}
\put(377.800,195.882){\line(0,1){3.086}}
\put(377.800,195.882){\line(1,0){9.100}}
\put(386.900,195.882){\line(0,1){3.086}}
\put(386.900,198.968){\line(1,0){9.100}}
\put(396.000,196.323){\line(0,1){2.646}}
\put(396.000,196.323){\line(1,0){9.100}}
\put(405.100,196.323){\line(0,1){2.646}}
\put(405.100,198.968){\line(1,0){9.100}}
\put(414.200,195.882){\line(0,1){3.086}}
\put(414.200,195.882){\line(1,0){9.100}}
\put(423.300,195.882){\line(0,1){3.527}}
\put(423.300,199.409){\line(1,0){9.100}}
\put(432.400,197.646){\line(0,1){1.764}}
\put(432.400,197.646){\line(1,0){9.100}}
\put(441.500,197.646){\line(0,1){2.646}}
\put(441.500,200.291){\line(1,0){9.100}}
\put(450.600,199.409){\line(0,1){0.882}}
\put(450.600,199.409){\line(1,0){9.100}}
\put(459.700,199.409){\line(0,1){3.086}}
\put(459.700,202.496){\line(1,0){9.100}}
\put(468.800,202.496){\line(0,1){14.109}}
\put(468.800,216.605){\line(1,0){9.100}}
\put(477.900,216.605){\line(0,1){46.296}}
\put(477.900,262.901){\line(1,0){9.100}}
\put(487.000,195.000){\line(0,1){67.901}}
\color[HTML]{BF5B35}
\linethickness{1pt}
\put(305.000,195.000){\line(0,1){9.259}}
\put(305.000,204.259){\line(1,0){9.100}}
\put(314.100,197.646){\line(0,1){6.614}}
\put(314.100,197.646){\line(1,0){9.100}}
\put(323.200,197.646){\line(0,1){0.000}}
\put(323.200,197.646){\line(1,0){9.100}}
\put(332.300,196.323){\line(0,1){1.323}}
\put(332.300,196.323){\line(1,0){9.100}}
\put(341.400,196.323){\line(0,1){0.882}}
\put(341.400,197.205){\line(1,0){9.100}}
\put(350.500,196.764){\line(0,1){0.441}}
\put(350.500,196.764){\line(1,0){9.100}}
\put(359.600,195.882){\line(0,1){0.882}}
\put(359.600,195.882){\line(1,0){9.100}}
\put(368.700,195.882){\line(0,1){0.441}}
\put(368.700,196.323){\line(1,0){9.100}}
\put(377.800,195.882){\line(0,1){0.441}}
\put(377.800,195.882){\line(1,0){9.100}}
\put(386.900,195.882){\line(0,1){1.323}}
\put(386.900,197.205){\line(1,0){9.100}}
\put(396.000,196.764){\line(0,1){0.441}}
\put(396.000,196.764){\line(1,0){9.100}}
\put(405.100,195.882){\line(0,1){0.882}}
\put(405.100,195.882){\line(1,0){9.100}}
\put(414.200,195.882){\line(0,1){2.205}}
\put(414.200,198.086){\line(1,0){9.100}}
\put(423.300,196.764){\line(0,1){1.323}}
\put(423.300,196.764){\line(1,0){9.100}}
\put(432.400,196.764){\line(0,1){0.441}}
\put(432.400,197.205){\line(1,0){9.100}}
\put(441.500,197.205){\line(0,1){1.323}}
\put(441.500,198.527){\line(1,0){9.100}}
\put(450.600,198.527){\line(0,1){2.646}}
\put(450.600,201.173){\line(1,0){9.100}}
\put(459.700,198.968){\line(0,1){2.205}}
\put(459.700,198.968){\line(1,0){9.100}}
\put(468.800,198.968){\line(0,1){4.409}}
\put(468.800,203.377){\line(1,0){9.100}}
\put(477.900,203.377){\line(0,1){84.215}}
\put(477.900,287.593){\line(1,0){9.100}}
\put(487.000,195.000){\line(0,1){92.593}}
\color{black}
\put(262.000,245.000){\rotatebox{90}{\small Count}}
\color{black}
\put(396.000,172.000){\makebox(0,0)[c]{\small Raw probability}}
\put(305.000,185.000){\makebox(0,0)[c]{\small 0}}
\put(396.000,185.000){\makebox(0,0)[c]{\small 0.5}}
\put(487.000,185.000){\makebox(0,0)[c]{\small 1}}
\put(300.000,195.000){\makebox(0,0)[r]{\small 0}}
\put(300.000,245.000){\makebox(0,0)[r]{\small 113}}
\put(300.000,295.000){\makebox(0,0)[r]{\small 227}}
\color{black}
\put(55.000,40.000){\line(1,0){182.000}}
\put(55.000,40.000){\line(0,1){100.000}}
\put(146.000,152.000){\makebox(0,0)[c]{\small Dense top-50; nonpositive}}
\color[HTML]{536878}
\linethickness{1pt}
\put(55.000,40.000){\line(0,1){64.727}}
\put(55.000,104.727){\line(1,0){9.100}}
\put(64.100,59.022){\line(0,1){45.705}}
\put(64.100,59.022){\line(1,0){9.100}}
\put(73.200,47.888){\line(0,1){11.134}}
\put(73.200,47.888){\line(1,0){9.100}}
\put(82.300,42.765){\line(0,1){5.122}}
\put(82.300,42.765){\line(1,0){9.100}}
\put(91.400,42.349){\line(0,1){0.416}}
\put(91.400,42.349){\line(1,0){9.100}}
\put(100.500,41.715){\line(0,1){0.635}}
\put(100.500,41.715){\line(1,0){9.100}}
\put(109.600,41.102){\line(0,1){0.613}}
\put(109.600,41.102){\line(1,0){9.100}}
\put(118.700,40.679){\line(0,1){0.423}}
\put(118.700,40.679){\line(1,0){9.100}}
\put(127.800,40.679){\line(0,1){0.197}}
\put(127.800,40.876){\line(1,0){9.100}}
\put(136.900,40.737){\line(0,1){0.139}}
\put(136.900,40.737){\line(1,0){9.100}}
\put(146.000,40.679){\line(0,1){0.058}}
\put(146.000,40.679){\line(1,0){9.100}}
\put(155.100,40.679){\line(0,1){0.109}}
\put(155.100,40.788){\line(1,0){9.100}}
\put(164.200,40.379){\line(0,1){0.409}}
\put(164.200,40.379){\line(1,0){9.100}}
\put(173.300,40.379){\line(0,1){0.241}}
\put(173.300,40.620){\line(1,0){9.100}}
\put(182.400,40.365){\line(0,1){0.255}}
\put(182.400,40.365){\line(1,0){9.100}}
\put(191.500,40.365){\line(0,1){0.073}}
\put(191.500,40.438){\line(1,0){9.100}}
\put(200.600,40.438){\line(0,1){0.139}}
\put(200.600,40.576){\line(1,0){9.100}}
\put(209.700,40.452){\line(0,1){0.124}}
\put(209.700,40.452){\line(1,0){9.100}}
\put(218.800,40.452){\line(0,1){0.277}}
\put(218.800,40.730){\line(1,0){9.100}}
\put(227.900,40.387){\line(0,1){0.343}}
\put(227.900,40.387){\line(1,0){9.100}}
\put(237.000,40.000){\line(0,1){0.387}}
\color[HTML]{BF5B35}
\linethickness{1pt}
\put(55.000,40.000){\line(0,1){92.593}}
\put(55.000,132.593){\line(1,0){9.100}}
\put(64.100,42.342){\line(0,1){90.250}}
\put(64.100,42.342){\line(1,0){9.100}}
\put(73.200,41.802){\line(0,1){0.540}}
\put(73.200,41.802){\line(1,0){9.100}}
\put(82.300,40.992){\line(0,1){0.810}}
\put(82.300,40.992){\line(1,0){9.100}}
\put(91.400,40.992){\line(0,1){0.015}}
\put(91.400,41.007){\line(1,0){9.100}}
\put(100.500,41.007){\line(0,1){0.051}}
\put(100.500,41.058){\line(1,0){9.100}}
\put(109.600,40.686){\line(0,1){0.372}}
\put(109.600,40.686){\line(1,0){9.100}}
\put(118.700,40.423){\line(0,1){0.263}}
\put(118.700,40.423){\line(1,0){9.100}}
\put(127.800,40.423){\line(0,1){0.051}}
\put(127.800,40.474){\line(1,0){9.100}}
\put(136.900,40.460){\line(0,1){0.015}}
\put(136.900,40.460){\line(1,0){9.100}}
\put(146.000,40.350){\line(0,1){0.109}}
\put(146.000,40.350){\line(1,0){9.100}}
\put(155.100,40.350){\line(0,1){0.022}}
\put(155.100,40.372){\line(1,0){9.100}}
\put(164.200,40.372){\line(0,1){0.000}}
\put(164.200,40.372){\line(1,0){9.100}}
\put(173.300,40.372){\line(0,1){0.146}}
\put(173.300,40.518){\line(1,0){9.100}}
\put(182.400,40.299){\line(0,1){0.219}}
\put(182.400,40.299){\line(1,0){9.100}}
\put(191.500,40.299){\line(0,1){0.233}}
\put(191.500,40.533){\line(1,0){9.100}}
\put(200.600,40.533){\line(0,1){0.029}}
\put(200.600,40.562){\line(1,0){9.100}}
\put(209.700,40.314){\line(0,1){0.248}}
\put(209.700,40.314){\line(1,0){9.100}}
\put(218.800,40.314){\line(0,1){0.029}}
\put(218.800,40.343){\line(1,0){9.100}}
\put(227.900,40.343){\line(0,1){1.430}}
\put(227.900,41.773){\line(1,0){9.100}}
\put(237.000,40.000){\line(0,1){1.773}}
\color{black}
\put(12.000,90.000){\rotatebox{90}{\small Count}}
\color{black}
\put(146.000,17.000){\makebox(0,0)[c]{\small Raw probability}}
\put(55.000,30.000){\makebox(0,0)[c]{\small 0}}
\put(146.000,30.000){\makebox(0,0)[c]{\small 0.5}}
\put(237.000,30.000){\makebox(0,0)[c]{\small 1}}
\put(50.000,40.000){\makebox(0,0)[r]{\small 0}}
\put(50.000,90.000){\makebox(0,0)[r]{\small 6853}}
\put(50.000,140.000){\makebox(0,0)[r]{\small 13705}}
\color{black}
\put(305.000,40.000){\line(1,0){182.000}}
\put(305.000,40.000){\line(0,1){100.000}}
\put(396.000,152.000){\makebox(0,0)[c]{\small Dense top-50; qrel positive}}
\color[HTML]{536878}
\linethickness{1pt}
\put(305.000,40.000){\line(0,1){0.452}}
\put(305.000,40.452){\line(1,0){9.100}}
\put(314.100,40.452){\line(0,1){0.000}}
\put(314.100,40.452){\line(1,0){9.100}}
\put(323.200,40.452){\line(0,1){0.452}}
\put(323.200,40.903){\line(1,0){9.100}}
\put(332.300,40.903){\line(0,1){0.903}}
\put(332.300,41.807){\line(1,0){9.100}}
\put(341.400,41.807){\line(0,1){0.903}}
\put(341.400,42.710){\line(1,0){9.100}}
\put(350.500,42.710){\line(0,1){0.452}}
\put(350.500,43.162){\line(1,0){9.100}}
\put(359.600,42.258){\line(0,1){0.903}}
\put(359.600,42.258){\line(1,0){9.100}}
\put(368.700,42.258){\line(0,1){0.452}}
\put(368.700,42.710){\line(1,0){9.100}}
\put(377.800,40.452){\line(0,1){2.258}}
\put(377.800,40.452){\line(1,0){9.100}}
\put(386.900,40.452){\line(0,1){1.807}}
\put(386.900,42.258){\line(1,0){9.100}}
\put(396.000,41.355){\line(0,1){0.903}}
\put(396.000,41.355){\line(1,0){9.100}}
\put(405.100,41.355){\line(0,1){1.807}}
\put(405.100,43.162){\line(1,0){9.100}}
\put(414.200,40.903){\line(0,1){2.258}}
\put(414.200,40.903){\line(1,0){9.100}}
\put(423.300,40.903){\line(0,1){2.710}}
\put(423.300,43.613){\line(1,0){9.100}}
\put(432.400,42.710){\line(0,1){0.903}}
\put(432.400,42.710){\line(1,0){9.100}}
\put(441.500,42.710){\line(0,1){0.903}}
\put(441.500,43.613){\line(1,0){9.100}}
\put(450.600,43.613){\line(0,1){0.903}}
\put(450.600,44.517){\line(1,0){9.100}}
\put(459.700,44.517){\line(0,1){2.710}}
\put(459.700,47.227){\line(1,0){9.100}}
\put(468.800,47.227){\line(0,1){14.453}}
\put(468.800,61.680){\line(1,0){9.100}}
\put(477.900,61.680){\line(0,1){46.974}}
\put(477.900,108.654){\line(1,0){9.100}}
\put(487.000,40.000){\line(0,1){68.654}}
\color[HTML]{BF5B35}
\linethickness{1pt}
\put(305.000,40.000){\line(0,1){4.065}}
\put(305.000,44.065){\line(1,0){9.100}}
\put(314.100,40.903){\line(0,1){3.162}}
\put(314.100,40.903){\line(1,0){9.100}}
\put(323.200,40.903){\line(0,1){0.452}}
\put(323.200,41.355){\line(1,0){9.100}}
\put(332.300,41.355){\line(0,1){0.000}}
\put(332.300,41.355){\line(1,0){9.100}}
\put(341.400,41.355){\line(0,1){0.000}}
\put(341.400,41.355){\line(1,0){9.100}}
\put(350.500,41.355){\line(0,1){0.452}}
\put(350.500,41.807){\line(1,0){9.100}}
\put(359.600,40.000){\line(0,1){1.807}}
\put(359.600,40.000){\line(1,0){9.100}}
\put(368.700,40.000){\line(0,1){1.355}}
\put(368.700,41.355){\line(1,0){9.100}}
\put(377.800,40.903){\line(0,1){0.452}}
\put(377.800,40.903){\line(1,0){9.100}}
\put(386.900,40.903){\line(0,1){0.903}}
\put(386.900,41.807){\line(1,0){9.100}}
\put(396.000,41.355){\line(0,1){0.452}}
\put(396.000,41.355){\line(1,0){9.100}}
\put(405.100,40.903){\line(0,1){0.452}}
\put(405.100,40.903){\line(1,0){9.100}}
\put(414.200,40.903){\line(0,1){0.903}}
\put(414.200,41.807){\line(1,0){9.100}}
\put(423.300,41.807){\line(0,1){0.000}}
\put(423.300,41.807){\line(1,0){9.100}}
\put(432.400,41.807){\line(0,1){0.000}}
\put(432.400,41.807){\line(1,0){9.100}}
\put(441.500,41.807){\line(0,1){0.903}}
\put(441.500,42.710){\line(1,0){9.100}}
\put(450.600,42.710){\line(0,1){2.710}}
\put(450.600,45.420){\line(1,0){9.100}}
\put(459.700,43.613){\line(0,1){1.807}}
\put(459.700,43.613){\line(1,0){9.100}}
\put(468.800,43.613){\line(0,1){4.065}}
\put(468.800,47.678){\line(1,0){9.100}}
\put(477.900,47.678){\line(0,1){84.914}}
\put(477.900,132.593){\line(1,0){9.100}}
\put(487.000,40.000){\line(0,1){92.593}}
\color{black}
\put(262.000,90.000){\rotatebox{90}{\small Count}}
\color{black}
\put(396.000,17.000){\makebox(0,0)[c]{\small Raw probability}}
\put(305.000,30.000){\makebox(0,0)[c]{\small 0}}
\put(396.000,30.000){\makebox(0,0)[c]{\small 0.5}}
\put(487.000,30.000){\makebox(0,0)[c]{\small 1}}
\put(300.000,40.000){\makebox(0,0)[r]{\small 0}}
\put(300.000,90.000){\makebox(0,0)[r]{\small 111}}
\put(300.000,140.000){\makebox(0,0)[r]{\small 221}}
\put(250.000,8.000){\makebox(0,0)[c]{\small \textcolor[HTML]{536878}{Jev}\quad\textcolor[HTML]{BF5B35}{OpenAI}}}
\end{picture}
}
\caption{Saved controlled-100 and dense Top-50 raw relevance scores, with shared bins and separate provider, qrel-status and candidate-condition panels. Nonpositive means no positive qrel. Histograms are descriptive. The decision interfaces do not guarantee interchangeable numerical scores.}\label{fig:scores}
\end{figure}
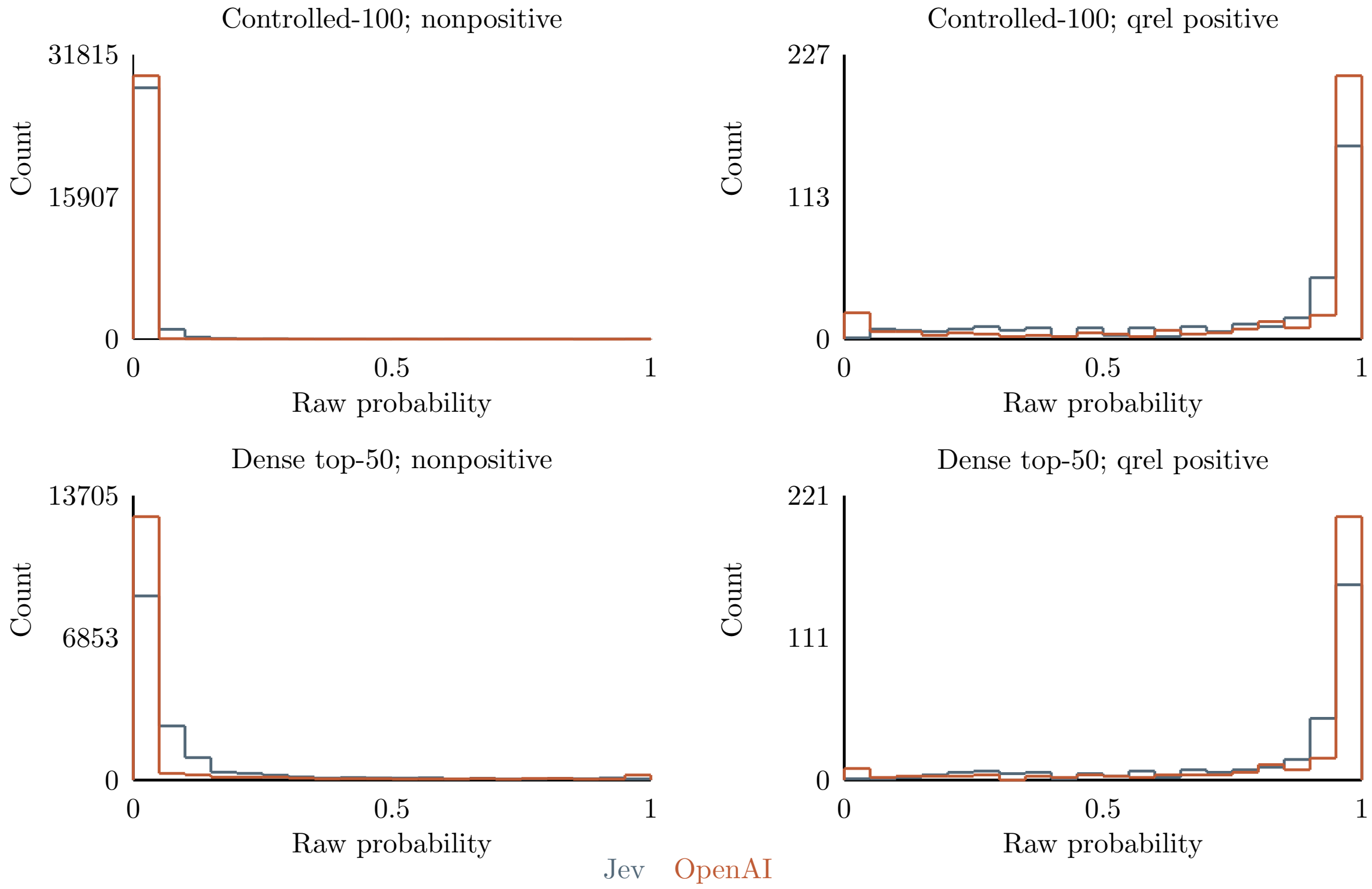

Figure~\ref{fig:scores} separates qrel-positive and nonpositive pairs within the controlled and Top-50 candidate conditions. Controlled positive mean scores are \JevControlledPositiveMean{} for Jev and \OpenAIControlledPositiveMean{} for OpenAI. Nonpositive means are \JevControlledNonpositiveMean{} and \OpenAIControlledNonpositiveMean{}. For hybrid candidates, positive means are \JevHybridPositiveMean{} and \OpenAIHybridPositiveMean{}, and nonpositive means are \JevHybridNonpositiveMean{} and \OpenAIHybridNonpositiveMean{}. Hybrid pools contain \OpenAIHybridPositiveCount{} known-positive pairs. The controlled pools contain \OpenAIControlledPositiveCount{}. Many high-scoring or low-scoring nonpositive pairs may be unjudged, so the plot does not establish perfect separation, verified irrelevance or calibration.

\subsection{Score saturation and tie sensitivity}
Table~\ref{tab:saturation} shows that OpenAI returns exact zero for \OpenAIControlledZero{} of controlled pairs and \OpenAIHybridZero{} of hybrid pairs, with exact-one fractions \OpenAIControlledOne{} and \OpenAIHybridOne{}. Jev returns neither endpoint in these records, but also has exact ties. Lack of endpoint saturation does not imply unique scores or calibration. OpenAI has \OpenAIControlledDistinct{} and \OpenAIHybridDistinct{} distinct returned values, respectively. We infer no internal numerical mechanism or training explanation.

Cutoff-crossing ties are common, but not all are mixed-label or consequential. OpenAI has \OpenAIControlledMixedBoundaryFive{} controlled and \OpenAIHybridMixedBoundaryFive{} hybrid queries with a mixed-label tie crossing the top-five cutoff. Corresponding top-ten counts are \OpenAIControlledMixedBoundaryTen{} and \OpenAIHybridMixedBoundaryTen{}. Mixed-label groups intersecting the top ten, rather than necessarily crossing its boundary, occur in \OpenAIControlledMixedTen{} and \OpenAIHybridMixedTen{} queries. A tied group wholly within the top ten can alter nDCG through position discounts without changing recall.

\begin{figure}[tb]\centering
\resizebox{\linewidth}{!}{%
\setlength{\unitlength}{1pt}
\begin{picture}(500,360)
\linethickness{0.6pt}
\color{black}
\put(55.000,225.000){\line(1,0){182.000}}
\put(55.000,225.000){\line(0,1){100.000}}
\put(146.000,337.000){\makebox(0,0)[c]{\small Controlled-100; Recall@5}}
\color{black}
\put(55.000,255.769){\line(1,0){4.000}}
\put(63.000,255.769){\line(1,0){4.000}}
\put(71.000,255.769){\line(1,0){4.000}}
\put(79.000,255.769){\line(1,0){4.000}}
\put(87.000,255.769){\line(1,0){4.000}}
\put(95.000,255.769){\line(1,0){4.000}}
\put(103.000,255.769){\line(1,0){4.000}}
\put(111.000,255.769){\line(1,0){4.000}}
\put(119.000,255.769){\line(1,0){4.000}}
\put(127.000,255.769){\line(1,0){4.000}}
\put(135.000,255.769){\line(1,0){4.000}}
\put(143.000,255.769){\line(1,0){4.000}}
\put(151.000,255.769){\line(1,0){4.000}}
\put(159.000,255.769){\line(1,0){4.000}}
\put(167.000,255.769){\line(1,0){4.000}}
\put(175.000,255.769){\line(1,0){4.000}}
\put(183.000,255.769){\line(1,0){4.000}}
\put(191.000,255.769){\line(1,0){4.000}}
\put(199.000,255.769){\line(1,0){4.000}}
\put(207.000,255.769){\line(1,0){4.000}}
\put(215.000,255.769){\line(1,0){4.000}}
\put(223.000,255.769){\line(1,0){4.000}}
\put(231.000,255.769){\line(1,0){4.000}}
\color[HTML]{536878}
\qbezier(55.000,236.538)(55.919,255.769)(56.838,275.000)
\qbezier(56.838,275.000)(57.758,255.769)(58.677,236.538)
\qbezier(58.677,236.538)(59.596,246.154)(60.515,255.769)
\qbezier(60.515,255.769)(61.434,246.154)(62.354,236.538)
\qbezier(62.354,236.538)(63.273,236.538)(64.192,236.538)
\qbezier(64.192,236.538)(65.111,236.538)(66.030,236.538)
\qbezier(66.030,236.538)(66.949,236.538)(67.869,236.538)
\qbezier(67.869,236.538)(68.788,236.538)(69.707,236.538)
\qbezier(69.707,236.538)(70.626,236.538)(71.545,236.538)
\qbezier(71.545,236.538)(72.465,246.154)(73.384,255.769)
\qbezier(73.384,255.769)(74.303,246.154)(75.222,236.538)
\qbezier(75.222,236.538)(76.141,275.000)(77.061,313.462)
\qbezier(77.061,313.462)(77.980,284.615)(78.899,255.769)
\qbezier(78.899,255.769)(79.818,246.154)(80.737,236.538)
\qbezier(80.737,236.538)(81.657,236.538)(82.576,236.538)
\qbezier(82.576,236.538)(83.495,236.538)(84.414,236.538)
\qbezier(84.414,236.538)(85.333,246.154)(86.253,255.769)
\qbezier(86.253,255.769)(87.172,255.769)(88.091,255.769)
\qbezier(88.091,255.769)(89.010,246.154)(89.929,236.538)
\qbezier(89.929,236.538)(90.848,236.538)(91.768,236.538)
\qbezier(91.768,236.538)(92.687,265.385)(93.606,294.231)
\qbezier(93.606,294.231)(94.525,284.615)(95.444,275.000)
\qbezier(95.444,275.000)(96.364,265.385)(97.283,255.769)
\qbezier(97.283,255.769)(98.202,246.154)(99.121,236.538)
\qbezier(99.121,236.538)(100.040,236.538)(100.960,236.538)
\qbezier(100.960,236.538)(101.879,255.769)(102.798,275.000)
\qbezier(102.798,275.000)(103.717,255.769)(104.636,236.538)
\qbezier(104.636,236.538)(105.556,236.538)(106.475,236.538)
\qbezier(106.475,236.538)(107.394,236.538)(108.313,236.538)
\qbezier(108.313,236.538)(109.232,236.538)(110.152,236.538)
\qbezier(110.152,236.538)(111.071,236.538)(111.990,236.538)
\qbezier(111.990,236.538)(112.909,236.538)(113.828,236.538)
\qbezier(113.828,236.538)(114.747,236.538)(115.667,236.538)
\qbezier(115.667,236.538)(116.586,246.154)(117.505,255.769)
\qbezier(117.505,255.769)(118.424,265.385)(119.343,275.000)
\qbezier(119.343,275.000)(120.263,255.769)(121.182,236.538)
\qbezier(121.182,236.538)(122.101,236.538)(123.020,236.538)
\qbezier(123.020,236.538)(123.939,236.538)(124.859,236.538)
\qbezier(124.859,236.538)(125.778,236.538)(126.697,236.538)
\qbezier(126.697,236.538)(127.616,236.538)(128.535,236.538)
\qbezier(128.535,236.538)(129.455,236.538)(130.374,236.538)
\qbezier(130.374,236.538)(131.293,236.538)(132.212,236.538)
\qbezier(132.212,236.538)(133.131,236.538)(134.051,236.538)
\qbezier(134.051,236.538)(134.970,236.538)(135.889,236.538)
\qbezier(135.889,236.538)(136.808,236.538)(137.727,236.538)
\qbezier(137.727,236.538)(138.646,236.538)(139.566,236.538)
\qbezier(139.566,236.538)(140.485,236.538)(141.404,236.538)
\qbezier(141.404,236.538)(142.323,236.538)(143.242,236.538)
\qbezier(143.242,236.538)(144.162,236.538)(145.081,236.538)
\qbezier(145.081,236.538)(146.000,236.538)(146.919,236.538)
\qbezier(146.919,236.538)(147.838,236.538)(148.758,236.538)
\qbezier(148.758,236.538)(149.677,236.538)(150.596,236.538)
\qbezier(150.596,236.538)(151.515,246.154)(152.434,255.769)
\qbezier(152.434,255.769)(153.354,246.154)(154.273,236.538)
\qbezier(154.273,236.538)(155.192,246.154)(156.111,255.769)
\qbezier(156.111,255.769)(157.030,246.154)(157.949,236.538)
\qbezier(157.949,236.538)(158.869,236.538)(159.788,236.538)
\qbezier(159.788,236.538)(160.707,236.538)(161.626,236.538)
\qbezier(161.626,236.538)(162.545,265.385)(163.465,294.231)
\qbezier(163.465,294.231)(164.384,275.000)(165.303,255.769)
\qbezier(165.303,255.769)(166.222,255.769)(167.141,255.769)
\qbezier(167.141,255.769)(168.061,265.385)(168.980,275.000)
\qbezier(168.980,275.000)(169.899,265.385)(170.818,255.769)
\qbezier(170.818,255.769)(171.737,255.769)(172.657,255.769)
\qbezier(172.657,255.769)(173.576,246.154)(174.495,236.538)
\qbezier(174.495,236.538)(175.414,236.538)(176.333,236.538)
\qbezier(176.333,236.538)(177.253,246.154)(178.172,255.769)
\qbezier(178.172,255.769)(179.091,246.154)(180.010,236.538)
\qbezier(180.010,236.538)(180.929,236.538)(181.848,236.538)
\qbezier(181.848,236.538)(182.768,236.538)(183.687,236.538)
\qbezier(183.687,236.538)(184.606,265.385)(185.525,294.231)
\qbezier(185.525,294.231)(186.444,275.000)(187.364,255.769)
\qbezier(187.364,255.769)(188.283,255.769)(189.202,255.769)
\qbezier(189.202,255.769)(190.121,255.769)(191.040,255.769)
\qbezier(191.040,255.769)(191.960,255.769)(192.879,255.769)
\qbezier(192.879,255.769)(193.798,246.154)(194.717,236.538)
\qbezier(194.717,236.538)(195.636,236.538)(196.556,236.538)
\qbezier(196.556,236.538)(197.475,236.538)(198.394,236.538)
\qbezier(198.394,236.538)(199.313,246.154)(200.232,255.769)
\qbezier(200.232,255.769)(201.152,246.154)(202.071,236.538)
\qbezier(202.071,236.538)(202.990,246.154)(203.909,255.769)
\qbezier(203.909,255.769)(204.828,246.154)(205.747,236.538)
\qbezier(205.747,236.538)(206.667,236.538)(207.586,236.538)
\qbezier(207.586,236.538)(208.505,236.538)(209.424,236.538)
\qbezier(209.424,236.538)(210.343,236.538)(211.263,236.538)
\qbezier(211.263,236.538)(212.182,236.538)(213.101,236.538)
\qbezier(213.101,236.538)(214.020,236.538)(214.939,236.538)
\qbezier(214.939,236.538)(215.859,236.538)(216.778,236.538)
\qbezier(216.778,236.538)(217.697,236.538)(218.616,236.538)
\qbezier(218.616,236.538)(219.535,246.154)(220.455,255.769)
\qbezier(220.455,255.769)(221.374,255.769)(222.293,255.769)
\qbezier(222.293,255.769)(223.212,246.154)(224.131,236.538)
\qbezier(224.131,236.538)(225.051,236.538)(225.970,236.538)
\qbezier(225.970,236.538)(226.889,246.154)(227.808,255.769)
\qbezier(227.808,255.769)(228.727,246.154)(229.646,236.538)
\qbezier(229.646,236.538)(230.566,236.538)(231.485,236.538)
\qbezier(231.485,236.538)(232.404,236.538)(233.323,236.538)
\qbezier(233.323,236.538)(234.242,236.538)(235.162,236.538)
\qbezier(235.162,236.538)(236.081,275.000)(237.000,313.462)
\color{black}
\put(146.000,202.000){\makebox(0,0)[c]{\small Predeclared tie seed}}
\put(55.000,215.000){\makebox(0,0)[c]{\small 0}}
\put(146.000,215.000){\makebox(0,0)[c]{\small 49.5}}
\put(237.000,215.000){\makebox(0,0)[c]{\small 99}}
\put(50.000,225.000){\makebox(0,0)[r]{\small 0.955}}
\put(50.000,275.000){\makebox(0,0)[r]{\small 0.963}}
\put(50.000,325.000){\makebox(0,0)[r]{\small 0.972}}
\color{black}
\put(305.000,225.000){\line(1,0){182.000}}
\put(305.000,225.000){\line(0,1){100.000}}
\put(396.000,337.000){\makebox(0,0)[c]{\small Controlled-100; nDCG@10}}
\color{black}
\put(305.000,261.606){\line(1,0){4.000}}
\put(313.000,261.606){\line(1,0){4.000}}
\put(321.000,261.606){\line(1,0){4.000}}
\put(329.000,261.606){\line(1,0){4.000}}
\put(337.000,261.606){\line(1,0){4.000}}
\put(345.000,261.606){\line(1,0){4.000}}
\put(353.000,261.606){\line(1,0){4.000}}
\put(361.000,261.606){\line(1,0){4.000}}
\put(369.000,261.606){\line(1,0){4.000}}
\put(377.000,261.606){\line(1,0){4.000}}
\put(385.000,261.606){\line(1,0){4.000}}
\put(393.000,261.606){\line(1,0){4.000}}
\put(401.000,261.606){\line(1,0){4.000}}
\put(409.000,261.606){\line(1,0){4.000}}
\put(417.000,261.606){\line(1,0){4.000}}
\put(425.000,261.606){\line(1,0){4.000}}
\put(433.000,261.606){\line(1,0){4.000}}
\put(441.000,261.606){\line(1,0){4.000}}
\put(449.000,261.606){\line(1,0){4.000}}
\put(457.000,261.606){\line(1,0){4.000}}
\put(465.000,261.606){\line(1,0){4.000}}
\put(473.000,261.606){\line(1,0){4.000}}
\put(481.000,261.606){\line(1,0){4.000}}
\color[HTML]{536878}
\qbezier(305.000,253.582)(305.919,255.616)(306.838,257.650)
\qbezier(306.838,257.650)(307.758,247.094)(308.677,236.538)
\qbezier(308.677,236.538)(309.596,244.699)(310.515,252.860)
\qbezier(310.515,252.860)(311.434,247.697)(312.354,242.534)
\qbezier(312.354,242.534)(313.273,248.058)(314.192,253.582)
\qbezier(314.192,253.582)(315.111,245.060)(316.030,236.538)
\qbezier(316.030,236.538)(316.949,245.060)(317.869,253.582)
\qbezier(317.869,253.582)(318.788,248.182)(319.707,242.783)
\qbezier(319.707,242.783)(320.626,251.180)(321.545,259.577)
\qbezier(321.545,259.577)(322.465,258.296)(323.384,257.014)
\qbezier(323.384,257.014)(324.303,261.877)(325.222,266.740)
\qbezier(325.222,266.740)(326.141,283.023)(327.061,299.306)
\qbezier(327.061,299.306)(327.980,282.987)(328.899,266.668)
\qbezier(328.899,266.668)(329.818,254.726)(330.737,242.783)
\qbezier(330.737,242.783)(331.657,239.660)(332.576,236.538)
\qbezier(332.576,236.538)(333.495,245.060)(334.414,253.582)
\qbezier(334.414,253.582)(335.333,261.889)(336.253,270.197)
\qbezier(336.253,270.197)(337.172,257.380)(338.091,244.563)
\qbezier(338.091,244.563)(339.010,240.551)(339.929,236.538)
\qbezier(339.929,236.538)(340.848,236.538)(341.768,236.538)
\qbezier(341.768,236.538)(342.687,252.015)(343.606,267.492)
\qbezier(343.606,267.492)(344.525,260.494)(345.444,253.496)
\qbezier(345.444,253.496)(346.364,257.551)(347.283,261.606)
\qbezier(347.283,261.606)(348.202,257.594)(349.121,253.582)
\qbezier(349.121,253.582)(350.040,253.582)(350.960,253.582)
\qbezier(350.960,253.582)(351.879,255.431)(352.798,257.281)
\qbezier(352.798,257.281)(353.717,255.431)(354.636,253.582)
\qbezier(354.636,253.582)(355.556,245.060)(356.475,236.538)
\qbezier(356.475,236.538)(357.394,236.538)(358.313,236.538)
\qbezier(358.313,236.538)(359.232,245.060)(360.152,253.582)
\qbezier(360.152,253.582)(361.071,248.182)(361.990,242.783)
\qbezier(361.990,242.783)(362.909,242.783)(363.828,242.783)
\qbezier(363.828,242.783)(364.747,239.660)(365.667,236.538)
\qbezier(365.667,236.538)(366.586,251.603)(367.505,266.668)
\qbezier(367.505,266.668)(368.424,268.604)(369.343,270.539)
\qbezier(369.343,270.539)(370.263,262.060)(371.182,253.582)
\qbezier(371.182,253.582)(372.101,245.060)(373.020,236.538)
\qbezier(373.020,236.538)(373.939,245.060)(374.859,253.582)
\qbezier(374.859,253.582)(375.778,245.060)(376.697,236.538)
\qbezier(376.697,236.538)(377.616,248.332)(378.535,260.125)
\qbezier(378.535,260.125)(379.455,256.853)(380.374,253.582)
\qbezier(380.374,253.582)(381.293,248.754)(382.212,243.927)
\qbezier(382.212,243.927)(383.131,240.233)(384.051,236.538)
\qbezier(384.051,236.538)(384.970,245.060)(385.889,253.582)
\qbezier(385.889,253.582)(386.808,259.851)(387.727,266.121)
\qbezier(387.727,266.121)(388.646,251.330)(389.566,236.538)
\qbezier(389.566,236.538)(390.485,245.060)(391.404,253.582)
\qbezier(391.404,253.582)(392.323,264.190)(393.242,274.798)
\qbezier(393.242,274.798)(394.162,259.363)(395.081,243.927)
\qbezier(395.081,243.927)(396.000,252.449)(396.919,260.970)
\qbezier(396.919,260.970)(397.838,251.752)(398.758,242.534)
\qbezier(398.758,242.534)(399.677,251.056)(400.596,259.577)
\qbezier(400.596,259.577)(401.515,256.515)(402.434,253.453)
\qbezier(402.434,253.453)(403.354,244.996)(404.273,236.538)
\qbezier(404.273,236.538)(405.192,241.005)(406.111,245.472)
\qbezier(406.111,245.472)(407.030,247.821)(407.949,250.171)
\qbezier(407.949,250.171)(408.869,252.597)(409.788,255.022)
\qbezier(409.788,255.022)(410.707,253.169)(411.626,251.316)
\qbezier(411.626,251.316)(412.545,272.376)(413.465,293.436)
\qbezier(413.465,293.436)(414.384,277.521)(415.303,261.606)
\qbezier(415.303,261.606)(416.222,261.606)(417.141,261.606)
\qbezier(417.141,261.606)(418.061,260.082)(418.980,258.559)
\qbezier(418.980,258.559)(419.899,261.256)(420.818,263.953)
\qbezier(420.818,263.953)(421.737,262.779)(422.657,261.606)
\qbezier(422.657,261.606)(423.576,257.594)(424.495,253.582)
\qbezier(424.495,253.582)(425.414,245.060)(426.333,236.538)
\qbezier(426.333,236.538)(427.253,241.724)(428.172,246.910)
\qbezier(428.172,246.910)(429.091,253.244)(430.010,259.577)
\qbezier(430.010,259.577)(430.929,248.058)(431.848,236.538)
\qbezier(431.848,236.538)(432.768,239.996)(433.687,243.453)
\qbezier(433.687,243.453)(434.606,261.727)(435.525,280.001)
\qbezier(435.525,280.001)(436.444,263.455)(437.364,246.910)
\qbezier(437.364,246.910)(438.283,264.973)(439.202,283.036)
\qbezier(439.202,283.036)(440.121,263.799)(441.040,244.563)
\qbezier(441.040,244.563)(441.960,255.616)(442.879,266.668)
\qbezier(442.879,266.668)(443.798,263.247)(444.717,259.826)
\qbezier(444.717,259.826)(445.636,256.704)(446.556,253.582)
\qbezier(446.556,253.582)(447.475,248.332)(448.394,243.082)
\qbezier(448.394,243.082)(449.313,253.453)(450.232,263.824)
\qbezier(450.232,263.824)(451.152,258.703)(452.071,253.582)
\qbezier(452.071,253.582)(452.990,258.048)(453.909,262.515)
\qbezier(453.909,262.515)(454.828,256.106)(455.747,249.697)
\qbezier(455.747,249.697)(456.667,254.911)(457.586,260.125)
\qbezier(457.586,260.125)(458.505,263.546)(459.424,266.966)
\qbezier(459.424,266.966)(460.343,258.481)(461.263,249.996)
\qbezier(461.263,249.996)(462.182,246.265)(463.101,242.534)
\qbezier(463.101,242.534)(464.020,242.808)(464.939,243.082)
\qbezier(464.939,243.082)(465.859,248.332)(466.778,253.582)
\qbezier(466.778,253.582)(467.697,251.454)(468.616,249.326)
\qbezier(468.616,249.326)(469.535,246.944)(470.455,244.563)
\qbezier(470.455,244.563)(471.374,249.008)(472.293,253.453)
\qbezier(472.293,253.453)(473.212,260.906)(474.131,268.359)
\qbezier(474.131,268.359)(475.051,260.970)(475.970,253.582)
\qbezier(475.970,253.582)(476.889,252.529)(477.808,251.477)
\qbezier(477.808,251.477)(478.727,244.008)(479.646,236.538)
\qbezier(479.646,236.538)(480.566,245.060)(481.485,253.582)
\qbezier(481.485,253.582)(482.404,245.060)(483.323,236.538)
\qbezier(483.323,236.538)(484.242,245.060)(485.162,253.582)
\qbezier(485.162,253.582)(486.081,283.522)(487.000,313.462)
\color{black}
\put(396.000,202.000){\makebox(0,0)[c]{\small Predeclared tie seed}}
\put(305.000,215.000){\makebox(0,0)[c]{\small 0}}
\put(396.000,215.000){\makebox(0,0)[c]{\small 49.5}}
\put(487.000,215.000){\makebox(0,0)[c]{\small 99}}
\put(300.000,225.000){\makebox(0,0)[r]{\small 0.940}}
\put(300.000,275.000){\makebox(0,0)[r]{\small 0.948}}
\put(300.000,325.000){\makebox(0,0)[r]{\small 0.956}}
\color{black}
\put(55.000,70.000){\line(1,0){182.000}}
\put(55.000,70.000){\line(0,1){100.000}}
\put(146.000,182.000){\makebox(0,0)[c]{\small Dense top-50; Recall@5}}
\color{black}
\put(55.000,141.739){\line(1,0){4.000}}
\put(63.000,141.739){\line(1,0){4.000}}
\put(71.000,141.739){\line(1,0){4.000}}
\put(79.000,141.739){\line(1,0){4.000}}
\put(87.000,141.739){\line(1,0){4.000}}
\put(95.000,141.739){\line(1,0){4.000}}
\put(103.000,141.739){\line(1,0){4.000}}
\put(111.000,141.739){\line(1,0){4.000}}
\put(119.000,141.739){\line(1,0){4.000}}
\put(127.000,141.739){\line(1,0){4.000}}
\put(135.000,141.739){\line(1,0){4.000}}
\put(143.000,141.739){\line(1,0){4.000}}
\put(151.000,141.739){\line(1,0){4.000}}
\put(159.000,141.739){\line(1,0){4.000}}
\put(167.000,141.739){\line(1,0){4.000}}
\put(175.000,141.739){\line(1,0){4.000}}
\put(183.000,141.739){\line(1,0){4.000}}
\put(191.000,141.739){\line(1,0){4.000}}
\put(199.000,141.739){\line(1,0){4.000}}
\put(207.000,141.739){\line(1,0){4.000}}
\put(215.000,141.739){\line(1,0){4.000}}
\put(223.000,141.739){\line(1,0){4.000}}
\put(231.000,141.739){\line(1,0){4.000}}
\color[HTML]{536878}
\qbezier(55.000,126.132)(55.919,114.426)(56.838,102.720)
\qbezier(56.838,102.720)(57.758,102.163)(58.677,101.605)
\qbezier(58.677,101.605)(59.596,118.328)(60.515,135.050)
\qbezier(60.515,135.050)(61.434,127.246)(62.354,119.443)
\qbezier(62.354,119.443)(63.273,125.574)(64.192,131.706)
\qbezier(64.192,131.706)(65.111,130.591)(66.030,129.476)
\qbezier(66.030,129.476)(66.949,125.017)(67.869,120.557)
\qbezier(67.869,120.557)(68.788,111.639)(69.707,102.720)
\qbezier(69.707,102.720)(70.626,103.278)(71.545,103.835)
\qbezier(71.545,103.835)(72.465,109.409)(73.384,114.983)
\qbezier(73.384,114.983)(74.303,127.804)(75.222,140.624)
\qbezier(75.222,140.624)(76.141,124.459)(77.061,108.294)
\qbezier(77.061,108.294)(77.980,108.294)(78.899,108.294)
\qbezier(78.899,108.294)(79.818,104.950)(80.737,101.605)
\qbezier(80.737,101.605)(81.657,107.737)(82.576,113.868)
\qbezier(82.576,113.868)(83.495,124.459)(84.414,135.050)
\qbezier(84.414,135.050)(85.333,124.459)(86.253,113.868)
\qbezier(86.253,113.868)(87.172,107.179)(88.091,100.491)
\qbezier(88.091,100.491)(89.010,101.605)(89.929,102.720)
\qbezier(89.929,102.720)(90.848,108.294)(91.768,113.868)
\qbezier(91.768,113.868)(92.687,108.294)(93.606,102.720)
\qbezier(93.606,102.720)(94.525,99.933)(95.444,97.146)
\qbezier(95.444,97.146)(96.364,102.720)(97.283,108.294)
\qbezier(97.283,108.294)(98.202,116.098)(99.121,123.902)
\qbezier(99.121,123.902)(100.040,132.821)(100.960,141.739)
\qbezier(100.960,141.739)(101.879,147.313)(102.798,152.887)
\qbezier(102.798,152.887)(103.717,144.526)(104.636,136.165)
\qbezier(104.636,136.165)(105.556,118.885)(106.475,101.605)
\qbezier(106.475,101.605)(107.394,97.146)(108.313,92.687)
\qbezier(108.313,92.687)(109.232,123.344)(110.152,154.002)
\qbezier(110.152,154.002)(111.071,122.787)(111.990,91.572)
\qbezier(111.990,91.572)(112.909,96.589)(113.828,101.605)
\qbezier(113.828,101.605)(114.747,118.328)(115.667,135.050)
\qbezier(115.667,135.050)(116.586,133.378)(117.505,131.706)
\qbezier(117.505,131.706)(118.424,136.165)(119.343,140.624)
\qbezier(119.343,140.624)(120.263,127.804)(121.182,114.983)
\qbezier(121.182,114.983)(122.101,114.426)(123.020,113.868)
\qbezier(123.020,113.868)(123.939,113.868)(124.859,113.868)
\qbezier(124.859,113.868)(125.778,106.065)(126.697,98.261)
\qbezier(126.697,98.261)(127.616,108.294)(128.535,118.328)
\qbezier(128.535,118.328)(129.455,121.672)(130.374,125.017)
\qbezier(130.374,125.017)(131.293,108.852)(132.212,92.687)
\qbezier(132.212,92.687)(133.131,103.278)(134.051,113.868)
\qbezier(134.051,113.868)(134.970,116.098)(135.889,118.328)
\qbezier(135.889,118.328)(136.808,122.230)(137.727,126.132)
\qbezier(137.727,126.132)(138.646,103.835)(139.566,81.538)
\qbezier(139.566,81.538)(140.485,95.474)(141.404,109.409)
\qbezier(141.404,109.409)(142.323,133.935)(143.242,158.462)
\qbezier(143.242,158.462)(144.162,120.000)(145.081,81.538)
\qbezier(145.081,81.538)(146.000,106.065)(146.919,130.591)
\qbezier(146.919,130.591)(147.838,121.672)(148.758,112.754)
\qbezier(148.758,112.754)(149.677,113.868)(150.596,114.983)
\qbezier(150.596,114.983)(151.515,125.574)(152.434,136.165)
\qbezier(152.434,136.165)(153.354,130.591)(154.273,125.017)
\qbezier(154.273,125.017)(155.192,113.311)(156.111,101.605)
\qbezier(156.111,101.605)(157.030,113.868)(157.949,126.132)
\qbezier(157.949,126.132)(158.869,117.770)(159.788,109.409)
\qbezier(159.788,109.409)(160.707,108.852)(161.626,108.294)
\qbezier(161.626,108.294)(162.545,102.163)(163.465,96.031)
\qbezier(163.465,96.031)(164.384,104.392)(165.303,112.754)
\qbezier(165.303,112.754)(166.222,123.902)(167.141,135.050)
\qbezier(167.141,135.050)(168.061,121.115)(168.980,107.179)
\qbezier(168.980,107.179)(169.899,129.476)(170.818,151.773)
\qbezier(170.818,151.773)(171.737,122.230)(172.657,92.687)
\qbezier(172.657,92.687)(173.576,122.787)(174.495,152.887)
\qbezier(174.495,152.887)(175.414,119.443)(176.333,85.998)
\qbezier(176.333,85.998)(177.253,88.227)(178.172,90.457)
\qbezier(178.172,90.457)(179.091,104.950)(180.010,119.443)
\qbezier(180.010,119.443)(180.929,119.443)(181.848,119.443)
\qbezier(181.848,119.443)(182.768,107.737)(183.687,96.031)
\qbezier(183.687,96.031)(184.606,104.950)(185.525,113.868)
\qbezier(185.525,113.868)(186.444,105.507)(187.364,97.146)
\qbezier(187.364,97.146)(188.283,99.376)(189.202,101.605)
\qbezier(189.202,101.605)(190.121,104.392)(191.040,107.179)
\qbezier(191.040,107.179)(191.960,116.656)(192.879,126.132)
\qbezier(192.879,126.132)(193.798,128.361)(194.717,130.591)
\qbezier(194.717,130.591)(195.636,125.574)(196.556,120.557)
\qbezier(196.556,120.557)(197.475,106.622)(198.394,92.687)
\qbezier(198.394,92.687)(199.313,97.703)(200.232,102.720)
\qbezier(200.232,102.720)(201.152,116.656)(202.071,130.591)
\qbezier(202.071,130.591)(202.990,131.148)(203.909,131.706)
\qbezier(203.909,131.706)(204.828,117.213)(205.747,102.720)
\qbezier(205.747,102.720)(206.667,121.115)(207.586,139.509)
\qbezier(207.586,139.509)(208.505,137.837)(209.424,136.165)
\qbezier(209.424,136.165)(210.343,122.230)(211.263,108.294)
\qbezier(211.263,108.294)(212.182,111.081)(213.101,113.868)
\qbezier(213.101,113.868)(214.020,110.524)(214.939,107.179)
\qbezier(214.939,107.179)(215.859,127.246)(216.778,147.313)
\qbezier(216.778,147.313)(217.697,122.230)(218.616,97.146)
\qbezier(218.616,97.146)(219.535,99.933)(220.455,102.720)
\qbezier(220.455,102.720)(221.374,102.163)(222.293,101.605)
\qbezier(222.293,101.605)(223.212,116.656)(224.131,131.706)
\qbezier(224.131,131.706)(225.051,122.787)(225.970,113.868)
\qbezier(225.970,113.868)(226.889,113.311)(227.808,112.754)
\qbezier(227.808,112.754)(228.727,121.115)(229.646,129.476)
\qbezier(229.646,129.476)(230.566,127.804)(231.485,126.132)
\qbezier(231.485,126.132)(232.404,134.493)(233.323,142.854)
\qbezier(233.323,142.854)(234.242,136.722)(235.162,130.591)
\qbezier(235.162,130.591)(236.081,116.656)(237.000,102.720)
\color{black}
\put(146.000,47.000){\makebox(0,0)[c]{\small Predeclared tie seed}}
\put(55.000,60.000){\makebox(0,0)[c]{\small 0}}
\put(146.000,60.000){\makebox(0,0)[c]{\small 49.5}}
\put(237.000,60.000){\makebox(0,0)[c]{\small 99}}
\put(50.000,70.000){\makebox(0,0)[r]{\small 0.792}}
\put(50.000,120.000){\makebox(0,0)[r]{\small 0.800}}
\put(50.000,170.000){\makebox(0,0)[r]{\small 0.807}}
\color{black}
\put(305.000,70.000){\line(1,0){182.000}}
\put(305.000,70.000){\line(0,1){100.000}}
\put(396.000,182.000){\makebox(0,0)[c]{\small Dense top-50; nDCG@10}}
\color{black}
\put(305.000,132.678){\line(1,0){4.000}}
\put(313.000,132.678){\line(1,0){4.000}}
\put(321.000,132.678){\line(1,0){4.000}}
\put(329.000,132.678){\line(1,0){4.000}}
\put(337.000,132.678){\line(1,0){4.000}}
\put(345.000,132.678){\line(1,0){4.000}}
\put(353.000,132.678){\line(1,0){4.000}}
\put(361.000,132.678){\line(1,0){4.000}}
\put(369.000,132.678){\line(1,0){4.000}}
\put(377.000,132.678){\line(1,0){4.000}}
\put(385.000,132.678){\line(1,0){4.000}}
\put(393.000,132.678){\line(1,0){4.000}}
\put(401.000,132.678){\line(1,0){4.000}}
\put(409.000,132.678){\line(1,0){4.000}}
\put(417.000,132.678){\line(1,0){4.000}}
\put(425.000,132.678){\line(1,0){4.000}}
\put(433.000,132.678){\line(1,0){4.000}}
\put(441.000,132.678){\line(1,0){4.000}}
\put(449.000,132.678){\line(1,0){4.000}}
\put(457.000,132.678){\line(1,0){4.000}}
\put(465.000,132.678){\line(1,0){4.000}}
\put(473.000,132.678){\line(1,0){4.000}}
\put(481.000,132.678){\line(1,0){4.000}}
\color[HTML]{536878}
\qbezier(305.000,84.329)(305.919,101.051)(306.838,117.773)
\qbezier(306.838,117.773)(307.758,123.237)(308.677,128.702)
\qbezier(308.677,128.702)(309.596,143.582)(310.515,158.462)
\qbezier(310.515,158.462)(311.434,133.757)(312.354,109.052)
\qbezier(312.354,109.052)(313.273,112.017)(314.192,114.982)
\qbezier(314.192,114.982)(315.111,117.609)(316.030,120.236)
\qbezier(316.030,120.236)(316.949,113.358)(317.869,106.481)
\qbezier(317.869,106.481)(318.788,101.672)(319.707,96.862)
\qbezier(319.707,96.862)(320.626,101.610)(321.545,106.358)
\qbezier(321.545,106.358)(322.465,112.478)(323.384,118.598)
\qbezier(323.384,118.598)(324.303,112.834)(325.222,107.070)
\qbezier(325.222,107.070)(326.141,104.665)(327.061,102.259)
\qbezier(327.061,102.259)(327.980,101.751)(328.899,101.243)
\qbezier(328.899,101.243)(329.818,113.054)(330.737,124.865)
\qbezier(330.737,124.865)(331.657,127.535)(332.576,130.205)
\qbezier(332.576,130.205)(333.495,125.222)(334.414,120.238)
\qbezier(334.414,120.238)(335.333,115.076)(336.253,109.914)
\qbezier(336.253,109.914)(337.172,118.857)(338.091,127.800)
\qbezier(338.091,127.800)(339.010,130.085)(339.929,132.371)
\qbezier(339.929,132.371)(340.848,129.532)(341.768,126.693)
\qbezier(341.768,126.693)(342.687,122.730)(343.606,118.767)
\qbezier(343.606,118.767)(344.525,115.151)(345.444,111.535)
\qbezier(345.444,111.535)(346.364,107.263)(347.283,102.992)
\qbezier(347.283,102.992)(348.202,117.851)(349.121,132.711)
\qbezier(349.121,132.711)(350.040,120.068)(350.960,107.426)
\qbezier(350.960,107.426)(351.879,124.458)(352.798,141.490)
\qbezier(352.798,141.490)(353.717,132.558)(354.636,123.626)
\qbezier(354.636,123.626)(355.556,123.837)(356.475,124.048)
\qbezier(356.475,124.048)(357.394,116.895)(358.313,109.742)
\qbezier(358.313,109.742)(359.232,129.283)(360.152,148.823)
\qbezier(360.152,148.823)(361.071,118.862)(361.990,88.900)
\qbezier(361.990,88.900)(362.909,99.017)(363.828,109.134)
\qbezier(363.828,109.134)(364.747,125.383)(365.667,141.633)
\qbezier(365.667,141.633)(366.586,133.647)(367.505,125.661)
\qbezier(367.505,125.661)(368.424,132.039)(369.343,138.416)
\qbezier(369.343,138.416)(370.263,123.629)(371.182,108.841)
\qbezier(371.182,108.841)(372.101,121.067)(373.020,133.293)
\qbezier(373.020,133.293)(373.939,119.229)(374.859,105.166)
\qbezier(374.859,105.166)(375.778,95.042)(376.697,84.918)
\qbezier(376.697,84.918)(377.616,91.605)(378.535,98.291)
\qbezier(378.535,98.291)(379.455,101.012)(380.374,103.732)
\qbezier(380.374,103.732)(381.293,110.905)(382.212,118.078)
\qbezier(382.212,118.078)(383.131,106.620)(384.051,95.162)
\qbezier(384.051,95.162)(384.970,99.063)(385.889,102.963)
\qbezier(385.889,102.963)(386.808,100.146)(387.727,97.330)
\qbezier(387.727,97.330)(388.646,89.674)(389.566,82.019)
\qbezier(389.566,82.019)(390.485,92.332)(391.404,102.645)
\qbezier(391.404,102.645)(392.323,119.648)(393.242,136.651)
\qbezier(393.242,136.651)(394.162,109.095)(395.081,81.538)
\qbezier(395.081,81.538)(396.000,108.154)(396.919,134.769)
\qbezier(396.919,134.769)(397.838,130.055)(398.758,125.341)
\qbezier(398.758,125.341)(399.677,113.788)(400.596,102.236)
\qbezier(400.596,102.236)(401.515,127.850)(402.434,153.465)
\qbezier(402.434,153.465)(403.354,142.671)(404.273,131.877)
\qbezier(404.273,131.877)(405.192,118.012)(406.111,104.147)
\qbezier(406.111,104.147)(407.030,110.521)(407.949,116.896)
\qbezier(407.949,116.896)(408.869,104.578)(409.788,92.260)
\qbezier(409.788,92.260)(410.707,106.393)(411.626,120.526)
\qbezier(411.626,120.526)(412.545,117.799)(413.465,115.072)
\qbezier(413.465,115.072)(414.384,108.099)(415.303,101.126)
\qbezier(415.303,101.126)(416.222,121.396)(417.141,141.666)
\qbezier(417.141,141.666)(418.061,126.513)(418.980,111.360)
\qbezier(418.980,111.360)(419.899,120.179)(420.818,128.998)
\qbezier(420.818,128.998)(421.737,119.285)(422.657,109.572)
\qbezier(422.657,109.572)(423.576,129.681)(424.495,149.790)
\qbezier(424.495,149.790)(425.414,120.219)(426.333,90.648)
\qbezier(426.333,90.648)(427.253,100.441)(428.172,110.235)
\qbezier(428.172,110.235)(429.091,120.743)(430.010,131.251)
\qbezier(430.010,131.251)(430.929,133.324)(431.848,135.396)
\qbezier(431.848,135.396)(432.768,115.710)(433.687,96.023)
\qbezier(433.687,96.023)(434.606,103.392)(435.525,110.761)
\qbezier(435.525,110.761)(436.444,112.582)(437.364,114.404)
\qbezier(437.364,114.404)(438.283,108.314)(439.202,102.223)
\qbezier(439.202,102.223)(440.121,106.850)(441.040,111.476)
\qbezier(441.040,111.476)(441.960,105.883)(442.879,100.289)
\qbezier(442.879,100.289)(443.798,102.115)(444.717,103.942)
\qbezier(444.717,103.942)(445.636,102.731)(446.556,101.521)
\qbezier(446.556,101.521)(447.475,101.514)(448.394,101.508)
\qbezier(448.394,101.508)(449.313,94.750)(450.232,87.992)
\qbezier(450.232,87.992)(451.152,103.844)(452.071,119.696)
\qbezier(452.071,119.696)(452.990,113.071)(453.909,106.447)
\qbezier(453.909,106.447)(454.828,98.084)(455.747,89.722)
\qbezier(455.747,89.722)(456.667,100.183)(457.586,110.644)
\qbezier(457.586,110.644)(458.505,113.946)(459.424,117.249)
\qbezier(459.424,117.249)(460.343,113.086)(461.263,108.923)
\qbezier(461.263,108.923)(462.182,117.572)(463.101,126.221)
\qbezier(463.101,126.221)(464.020,119.782)(464.939,113.343)
\qbezier(464.939,113.343)(465.859,106.420)(466.778,99.498)
\qbezier(466.778,99.498)(467.697,99.068)(468.616,98.638)
\qbezier(468.616,98.638)(469.535,101.574)(470.455,104.509)
\qbezier(470.455,104.509)(471.374,112.967)(472.293,121.425)
\qbezier(472.293,121.425)(473.212,128.309)(474.131,135.192)
\qbezier(474.131,135.192)(475.051,120.057)(475.970,104.921)
\qbezier(475.970,104.921)(476.889,107.176)(477.808,109.431)
\qbezier(477.808,109.431)(478.727,112.372)(479.646,115.313)
\qbezier(479.646,115.313)(480.566,112.581)(481.485,109.849)
\qbezier(481.485,109.849)(482.404,105.138)(483.323,100.427)
\qbezier(483.323,100.427)(484.242,108.593)(485.162,116.760)
\qbezier(485.162,116.760)(486.081,101.262)(487.000,85.765)
\color{black}
\put(396.000,47.000){\makebox(0,0)[c]{\small Predeclared tie seed}}
\put(305.000,60.000){\makebox(0,0)[c]{\small 0}}
\put(396.000,60.000){\makebox(0,0)[c]{\small 49.5}}
\put(487.000,60.000){\makebox(0,0)[c]{\small 99}}
\put(300.000,70.000){\makebox(0,0)[r]{\small 0.746}}
\put(300.000,120.000){\makebox(0,0)[r]{\small 0.753}}
\put(300.000,170.000){\makebox(0,0)[r]{\small 0.759}}
\color[HTML]{536878}
\put(40.000,14.000){\line(1,0){25.000}}
\put(72.000,14.000){\makebox(0,0)[l]{\small Exact-tie realizations}}
\color{black}
\put(270.000,14.000){\line(1,0){8.000}}
\put(284.000,14.000){\line(1,0){8.000}}
\put(302.000,14.000){\makebox(0,0)[l]{\small Primary document-ID rule}}
\end{picture}
}
\caption{OpenAI controlled-100 and dense Top-50 secondary exact-tie realizations for all seeds 0--99. Black reference lines are the primary document-ID policy. Each panel has its own scale. This is sensitivity within a protocol, not a cross-condition discovery comparison.}\label{fig:ties}
\end{figure}
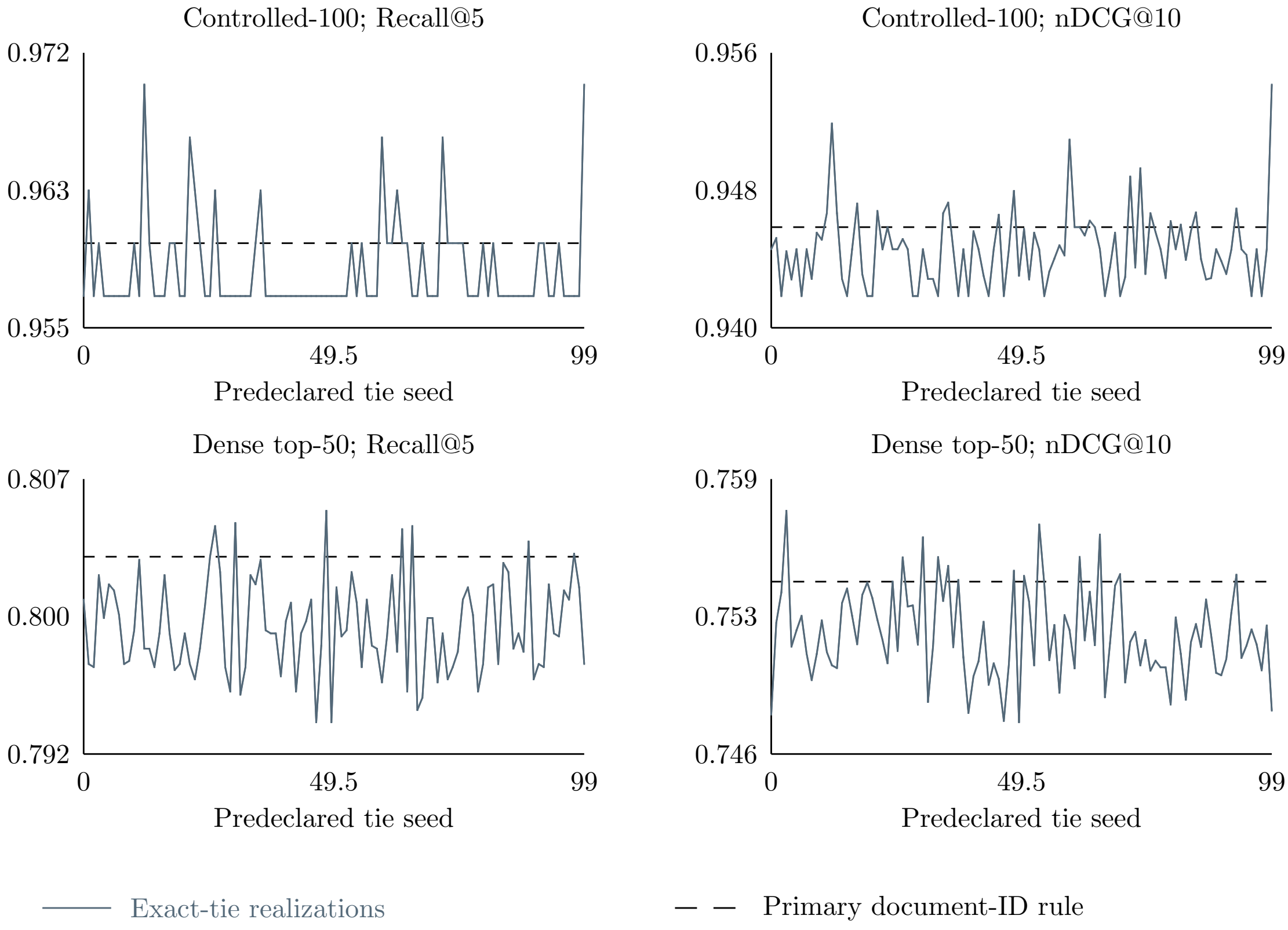

Table~\ref{tab:ties} and Figure~\ref{fig:ties} report the secondary exact-tie analysis. Controlled Recall@5 changes for \ControlledOpenAITieRecallAffected{} queries across the realizations. Controlled nDCG@10 changes for \ControlledOpenAITieNDCGAffected{}. Hybrid counts are \HybridOpenAITieRecallAffected{} and \HybridOpenAITieNDCGAffected{}. These are observed sensitivity ranges over 100 seeded realizations, not theoretical permutation bounds. The original deterministic rankings and primary metrics remain unchanged.
\begin{table}[tb]\centering\footnotesize
\caption{Raw-score saturation and queries with cutoff-crossing or mixed-label ties. A mixed tie intersecting the top ten need not cross its cutoff or change a metric.}\label{tab:saturation}
\begin{tabular}{@{}lrrrrrr@{}}\toprule
Condition & $p=0$ & $p=1$ & Distinct & Cutoff 5 & Cutoff 10 & Mixed top 10\\\midrule
Jev controlled & \JevControlledZero{} & \JevControlledOne{} & \JevControlledDistinct{} & \JevControlledBoundaryFive{} & \JevControlledBoundaryTen{} & \JevControlledMixedTen{}\\
OpenAI controlled & \OpenAIControlledZero{} & \OpenAIControlledOne{} & \OpenAIControlledDistinct{} & \OpenAIControlledBoundaryFive{} & \OpenAIControlledBoundaryTen{} & \OpenAIControlledMixedTen{}\\
Jev hybrid & \JevHybridZero{} & \JevHybridOne{} & \JevHybridDistinct{} & \JevHybridBoundaryFive{} & \JevHybridBoundaryTen{} & \JevHybridMixedTen{}\\
OpenAI hybrid & \OpenAIHybridZero{} & \OpenAIHybridOne{} & \OpenAIHybridDistinct{} & \OpenAIHybridBoundaryFive{} & \OpenAIHybridBoundaryTen{} & \OpenAIHybridMixedTen{}\\
\bottomrule\end{tabular}\end{table}

\begin{table}[tb]\centering\footnotesize
\caption{OpenAI exact-tie sensitivity over seeds 0--99. Ranges cover 100 seeded realizations, not exhaustive permutation bounds. Changed queries differ from the primary policy in at least one realization.}\label{tab:ties}
\begin{tabular}{@{}llrrrr@{}}\toprule
Condition & Metric & Primary & Seed mean & Seed range & Changed queries\\\midrule
Controlled & Recall@5 & \ControlledOpenAITieRecallPrimary{} & \ControlledOpenAITieRecallMean{} & $[\ControlledOpenAITieRecallMin{}, \ControlledOpenAITieRecallMax{}]$ & \ControlledOpenAITieRecallAffected{}\\
Controlled & nDCG@10 & \ControlledOpenAITieNDCGPrimary{} & \ControlledOpenAITieNDCGMean{} & $[\ControlledOpenAITieNDCGMin{}, \ControlledOpenAITieNDCGMax{}]$ & \ControlledOpenAITieNDCGAffected{}\\
Hybrid & Recall@5 & \HybridOpenAITieRecallPrimary{} & \HybridOpenAITieRecallMean{} & $[\HybridOpenAITieRecallMin{}, \HybridOpenAITieRecallMax{}]$ & \HybridOpenAITieRecallAffected{}\\
Hybrid & nDCG@10 & \HybridOpenAITieNDCGPrimary{} & \HybridOpenAITieNDCGMean{} & $[\HybridOpenAITieNDCGMin{}, \HybridOpenAITieNDCGMax{}]$ & \HybridOpenAITieNDCGAffected{}\\
\bottomrule\end{tabular}\end{table}

\subsection{Usage-derived cost and operational latency}
Table~\ref{tab:usage} reports successful-decision usage and unique condition cost, including reused decisions once. Standard-rate USD/query is \ControlledQueryCost{} for controlled Jev, \ControlledOpenAIQueryCost{} for controlled OpenAI, \HybridQueryCost{} for hybrid Jev, and \HybridOpenAIQueryCost{} for hybrid OpenAI. The recorded OpenAI controlled/Top-50 usage implies \ExtensionFreshCost{} USD of usage-derived fresh successful-call expenditure over \ExtensionFreshDecisions{} new unique pairs. This amount excludes historical reused decisions. Jev controlled/hybrid fresh successful-call estimates are \ControlledNewSpend{} and \HybridNewSpend{} USD. The OpenAI separate controlled/hybrid fresh estimates are \ControlledOpenAIFreshCost{} and \HybridOpenAIFreshCost{} USD. Cached pairs overlap, so unique condition costs must not simply be summed as fresh spend. Local BGE monetary cost was not measured.

Table~\ref{tab:latency} presents mean, median and p95 under the recorded timing definitions. These are operational observations, not intrinsic model speeds. BGE used local Apple MPS, whereas both decision APIs were hosted remotely. Their cache states, gateways, retries and timer boundaries differed. Hybrid totals sum saved dense timing and separately measured reranking and are not freshly measured combined replays. Controlled Jev includes reused pilot results and active repair components. Individual decision latency, scoring/persistence wall time, full ranking time and cache-only replay are retained separately and are not pooled into a common model-inference estimate.

\subsection{Candidate-Depth Sensitivity at Top-50 and Top-100}\label{sec:depthresults}
\paragraph{First-stage coverage and final effectiveness.}
Known-positive candidate recall rises from \DepthFifty{} at depth 50 to \DepthHundred{} at depth 100. Ranks 51--100 add \DepthAddedPairs{} known-positive pairs for \DepthGainingQueries{} queries. Already \DepthCompleteFifty{} queries contained every known positive at depth 50, while \DepthMissingHundred{} still miss at least one at depth 100. This is improved candidate availability, not a guarantee of improved final ranking. Table~\ref{tab:hybrid} includes all five full-corpus conditions. Jev's Top-100 Recall@5 and nDCG@10 are \DepthJevRecall{} and \DepthJevNDCG{}. OpenAI's are \DepthOpenAIRecall{} and \DepthOpenAINDCG{}. Jev's point estimates rise on both metrics. OpenAI's Recall rises while its nDCG falls slightly. These dense-prefix conditions must remain separate from positive-enriched Controlled-100.

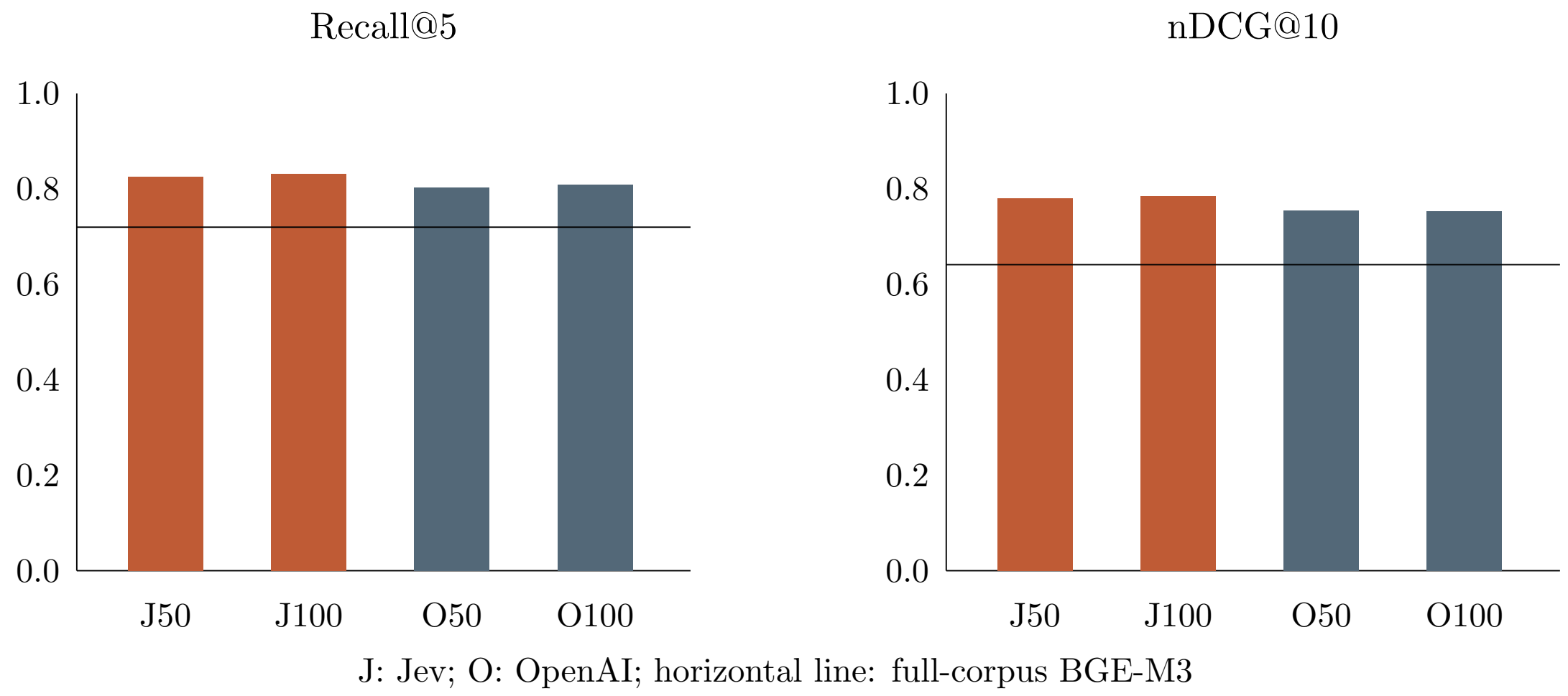
\begin{figure}[tb]\centering
\resizebox{\linewidth}{!}{%
\setlength{\unitlength}{1pt}
\begin{picture}(500,210)
\put(45.000,35.000){\line(1,0){180}}
\put(45.000,35.000){\line(0,1){140}}
\put(135.000,195.000){\makebox(0,0){Recall@5}}
\put(40.000,35.000){\makebox(0,0)[r]{\small 0.0}}
\put(40.000,63.000){\makebox(0,0)[r]{\small 0.2}}
\put(40.000,91.000){\makebox(0,0)[r]{\small 0.4}}
\put(40.000,119.000){\makebox(0,0)[r]{\small 0.6}}
\put(40.000,147.000){\makebox(0,0)[r]{\small 0.8}}
\put(40.000,175.000){\makebox(0,0)[r]{\small 1.0}}
\put(60.000,35.000){\textcolor[HTML]{BF5B35}{\rule{22pt}{115.656pt}}}
\put(71.000,22.000){\makebox(0,0){\small J50}}
\put(102.000,35.000){\textcolor[HTML]{BF5B35}{\rule{22pt}{116.356pt}}}
\put(113.000,22.000){\makebox(0,0){\small J100}}
\put(144.000,35.000){\textcolor[HTML]{536878}{\rule{22pt}{112.389pt}}}
\put(155.000,22.000){\makebox(0,0){\small O50}}
\put(186.000,35.000){\textcolor[HTML]{536878}{\rule{22pt}{113.229pt}}}
\put(197.000,22.000){\makebox(0,0){\small O100}}
\put(45.000,135.784){\line(1,0){180}}
\put(300.000,35.000){\line(1,0){180}}
\put(300.000,35.000){\line(0,1){140}}
\put(390.000,195.000){\makebox(0,0){nDCG@10}}
\put(295.000,35.000){\makebox(0,0)[r]{\small 0.0}}
\put(295.000,63.000){\makebox(0,0)[r]{\small 0.2}}
\put(295.000,91.000){\makebox(0,0)[r]{\small 0.4}}
\put(295.000,119.000){\makebox(0,0)[r]{\small 0.6}}
\put(295.000,147.000){\makebox(0,0)[r]{\small 0.8}}
\put(295.000,175.000){\makebox(0,0)[r]{\small 1.0}}
\put(315.000,35.000){\textcolor[HTML]{BF5B35}{\rule{22pt}{109.267pt}}}
\put(326.000,22.000){\makebox(0,0){\small J50}}
\put(357.000,35.000){\textcolor[HTML]{BF5B35}{\rule{22pt}{109.926pt}}}
\put(368.000,22.000){\makebox(0,0){\small J100}}
\put(399.000,35.000){\textcolor[HTML]{536878}{\rule{22pt}{105.626pt}}}
\put(410.000,22.000){\makebox(0,0){\small O50}}
\put(441.000,35.000){\textcolor[HTML]{536878}{\rule{22pt}{105.421pt}}}
\put(452.000,22.000){\makebox(0,0){\small O100}}
\put(300.000,124.779){\line(1,0){180}}
\put(250,5){\makebox(0,0){\small J: Jev; O: OpenAI; horizontal line: full-corpus BGE-M3}}
\end{picture}
}
\caption{Exploratory full-corpus dense-prefix effectiveness for Jev and OpenAI at exactly depths 50 and 100, without positive injection. Controlled-100 is excluded. The horizontal reference is the unchanged full-corpus dense baseline. Both bar-chart axes run from zero to one. Bars are point estimates. All four depth-change intervals in Table~\ref{tab:depth} include zero. No intermediate depths or extrapolation are shown.}\label{fig:depth}
\end{figure}

\paragraph{Paired depth analysis.}
Table~\ref{tab:depth} reports all four predefined depth contrasts. All four unadjusted 95\% intervals include zero. The evidence therefore supports small and uncertain final-metric changes, not a reliably positive depth effect. Including zero is not equivalence or proof of no possible effect. Table~\ref{tab:depthcross} reports additional Top-100 comparisons. Both providers' Top-100-minus-dense intervals are positive for both metrics. OpenAI-minus-Jev Top-100 Recall@5 includes zero, while its nDCG@10 interval is negative. This does not establish provider superiority on every metric. Original Top-50 inference remains separately reported.
\begin{table}[tbp]\centering\footnotesize
\caption{Exploratory depth differences, 100 minus 50, with paired percentile 95\% intervals, 10,000 resamples, seed 42. All four intervals include zero. Intervals are unadjusted. I/D/T denotes improved/declined/tied queries.}\label{tab:depth}
\begin{tabular}{@{}llrrl@{}}\toprule
Comparison & Metric & $\Delta$ & 95\% interval & I/D/T\\\midrule
Jev 100 minus Jev 50 & Recall@5 & \DepthContrastADelta{} & $[\DepthContrastALower{}, \DepthContrastAUpper{}]$ & \DepthContrastAImproved{}/\DepthContrastADeclined{}/\DepthContrastATied{}\\
Jev 100 minus Jev 50 & nDCG@10 & \DepthContrastBDelta{} & $[\DepthContrastBLower{}, \DepthContrastBUpper{}]$ & \DepthContrastBImproved{}/\DepthContrastBDeclined{}/\DepthContrastBTied{}\\
OpenAI 100 minus OpenAI 50 & Recall@5 & \DepthContrastCDelta{} & $[\DepthContrastCLower{}, \DepthContrastCUpper{}]$ & \DepthContrastCImproved{}/\DepthContrastCDeclined{}/\DepthContrastCTied{}\\
OpenAI 100 minus OpenAI 50 & nDCG@10 & \DepthContrastDDelta{} & $[\DepthContrastDLower{}, \DepthContrastDUpper{}]$ & \DepthContrastDImproved{}/\DepthContrastDDeclined{}/\DepthContrastDTied{}\\
\bottomrule\end{tabular}\end{table}

\paragraph{Ranking stability and score behavior.}
The relative order of every original Top-50 list is preserved for both providers across all \DatasetQueries{} queries. Newly added candidates can intrude into the final top five or ten and displace prior candidates without changing any historical pair score. Jev Top-100 has exact-zero and exact-one fractions \DepthJevZero{} and \DepthJevOne{}, \DepthJevDistinct{} distinct values and tied-pair fraction \DepthJevTiedPairs{}. OpenAI has corresponding fractions \DepthOpenAIZero{} and \DepthOpenAIOne{}, \DepthOpenAIDistinct{} values and tied-pair fraction \DepthOpenAITiedPairs{}. OpenAI again exhibits substantial zero saturation. Jev also produces ties.

Top-five/top-ten cutoff ties affect \DepthJevCutoffFive{}/\DepthJevCutoffTen{} Jev queries and \DepthOpenAICutoffFive{}/\DepthOpenAICutoffTen{} OpenAI queries. Mixed-label ties intersecting the top ten occur in \DepthJevMixedTen{} and \DepthOpenAIMixedTen{} queries. Mixed-label cutoff-crossing counts are \DepthJevImpactFive{}/\DepthJevImpactTen{} and \DepthOpenAIImpactFive{}/\DepthOpenAIImpactTen{}. Table~\ref{tab:depthties} retains primary metrics, seeded means and observed ranges. These ranges cover 100 realizations, not all possible permutations. Nonpositive documents may be unjudged, endpoint saturation is not calibration, and no internal model mechanism follows from these diagnostics.

\paragraph{Cost and incremental execution.}
Each provider's Top-100 condition contains twice the candidate decisions of Top-50. Both reuse \DepthJevCached{} historical compatible decisions and execute \DepthJevNew{} new ones. Table~\ref{tab:depthcost} separates unique-condition scoring cost from incremental new successful-call expenditure. Jev USD/query changes from \HybridQueryCost{} to \DepthJevQueryCost{}. OpenAI USD/query changes from \HybridOpenAIQueryCost{} to \DepthOpenAIQueryCost{}, approximately doubling unique-condition cost. Incremental Top-100 expenditure is \DepthJevIncremental{} USD for Jev and \DepthOpenAIIncremental{} for OpenAI. Cache reuse prevents interpreting this as doubling new expenditure. 
\begin{table}[tbp]\centering\footnotesize
\caption{Top-100 scoring accounting in USD. Unique cost includes each compatible decision once. Incremental cost counts only newly successful decisions.}\label{tab:depthcost}
\begin{tabular}{@{}lrrrrr@{}}\toprule
Provider & Cached & New & Unique USD & New USD & USD/query\\\midrule
Jev & \DepthJevCached{} & \DepthJevNew{} & \DepthJevCost{} & \DepthJevIncremental{} & \DepthJevQueryCost{}\\
OpenAI & \DepthOpenAICached{} & \DepthOpenAINew{} & \DepthOpenAICost{} & \DepthOpenAIIncremental{} & \DepthOpenAIQueryCost{}\\
\bottomrule\end{tabular}\end{table}

\begin{samepage}
The mixed execution cache-hit fraction is \DepthJevCacheFraction{} for each provider. Query medians are \DepthJevMixedMedian{} seconds for Jev and \DepthOpenAIMixedMedian{} for OpenAI. Table~\ref{tab:depthlatency} separates the saved timers. These observations do not measure fresh cold scoring of 100 candidates, and historical Top-50 timers have different boundaries. They cannot establish a relative cold-serving speed or a percentage latency change from doubling depth.\par
\end{samepage}

\subsection{Additional Top-100 comparisons and operational diagnostics}
The additional provider and dense-baseline contrasts, Top-100 tie realizations and separately defined timers are retained in Tables~\ref{tab:depthcross}--\ref{tab:depthlatency}. These supplement the primary depth comparisons without changing the deterministic rankings or pooling distinct timing boundaries. The confidence intervals remain exploratory and unadjusted. The tie ranges cover only the specified seed realizations.
\begin{table}[htbp]\centering\footnotesize
\caption{Additional exploratory Top-100 contrasts; the Top-50 inference remains separate. Unadjusted query-paired percentile intervals.}\label{tab:depthcross}
\begin{tabular}{@{}llrrl@{}}\toprule
Comparison & Metric & $\Delta$ & 95\% interval & I/D/T\\\midrule
Jev 100 minus BGE & Recall@5 & \DepthContrastEDelta{} & $[\DepthContrastELower{}, \DepthContrastEUpper{}]$ & \DepthContrastEImproved{}/\DepthContrastEDeclined{}/\DepthContrastETied{}\\
Jev 100 minus BGE & nDCG@10 & \DepthContrastFDelta{} & $[\DepthContrastFLower{}, \DepthContrastFUpper{}]$ & \DepthContrastFImproved{}/\DepthContrastFDeclined{}/\DepthContrastFTied{}\\
OpenAI 100 minus BGE & Recall@5 & \DepthContrastGDelta{} & $[\DepthContrastGLower{}, \DepthContrastGUpper{}]$ & \DepthContrastGImproved{}/\DepthContrastGDeclined{}/\DepthContrastGTied{}\\
OpenAI 100 minus BGE & nDCG@10 & \DepthContrastHDelta{} & $[\DepthContrastHLower{}, \DepthContrastHUpper{}]$ & \DepthContrastHImproved{}/\DepthContrastHDeclined{}/\DepthContrastHTied{}\\
OpenAI 100 minus Jev 100 & Recall@5 & \DepthContrastIDelta{} & $[\DepthContrastILower{}, \DepthContrastIUpper{}]$ & \DepthContrastIImproved{}/\DepthContrastIDeclined{}/\DepthContrastITied{}\\
OpenAI 100 minus Jev 100 & nDCG@10 & \DepthContrastJDelta{} & $[\DepthContrastJLower{}, \DepthContrastJUpper{}]$ & \DepthContrastJImproved{}/\DepthContrastJDeclined{}/\DepthContrastJTied{}\\
\bottomrule\end{tabular}\end{table}

\begin{table}[htbp]\centering\footnotesize
\caption{Dense Top-100 exact-tie sensitivity over seeds 0--99. Observed ranges are not exhaustive permutation bounds. Official document-ID rankings remain unchanged.}\label{tab:depthties}
\begin{tabular}{@{}llrrrr@{}}\toprule
Provider & Metric & Primary & Seed mean & Seed range & Changed\\\midrule
Jev & Recall@5 & \DepthJevTieRecallPrimary{} & \DepthJevTieRecallMean{} & $[\DepthJevTieRecallMin{}, \DepthJevTieRecallMax{}]$ & \DepthJevTieRecallChanged{}\\
Jev & nDCG@10 & \DepthJevTieNDCGPrimary{} & \DepthJevTieNDCGMean{} & $[\DepthJevTieNDCGMin{}, \DepthJevTieNDCGMax{}]$ & \DepthJevTieNDCGChanged{}\\
OpenAI & Recall@5 & \DepthOpenAITieRecallPrimary{} & \DepthOpenAITieRecallMean{} & $[\DepthOpenAITieRecallMin{}, \DepthOpenAITieRecallMax{}]$ & \DepthOpenAITieRecallChanged{}\\
OpenAI & nDCG@10 & \DepthOpenAITieNDCGPrimary{} & \DepthOpenAITieNDCGMean{} & $[\DepthOpenAITieNDCGMin{}, \DepthOpenAITieNDCGMax{}]$ & \DepthOpenAITieNDCGChanged{}\\
\bottomrule\end{tabular}\end{table}

\begin{table}[htbp]\centering\footnotesize
\caption{Top-100 observed seconds from separate saved timers, mixed cached/new execution and cache-only replay. No cold 100-document inference or combined dense-plus-API timing was measured. Historical Top-50 timers have different boundaries.}\label{tab:depthlatency}
\begin{tabular}{@{}llrrr@{}}\toprule
Provider & Timer & Mean & Median & p95\\\midrule
Jev & New HTTP request & \DepthJevRequestMean{} & \DepthJevRequestMedian{} & \DepthJevRequestPninetyfive{}\\
Jev & Incremental scoring & \DepthJevScoringMean{} & \DepthJevScoringMedian{} & \DepthJevScoringPninetyfive{}\\
Jev & Cache lookup & \DepthJevCacheMean{} & \DepthJevCacheMedian{} & \DepthJevCachePninetyfive{}\\
Jev & Ranking assembly & \DepthJevAssemblyMean{} & \DepthJevAssemblyMedian{} & \DepthJevAssemblyPninetyfive{}\\
Jev & Mixed query wall & \DepthJevMixedMean{} & \DepthJevMixedMedian{} & \DepthJevMixedPninetyfive{}\\
Jev & Cache-only replay & \DepthJevReplayMean{} & \DepthJevReplayMedian{} & \DepthJevReplayPninetyfive{}\\
OpenAI & New HTTP request & \DepthOpenAIRequestMean{} & \DepthOpenAIRequestMedian{} & \DepthOpenAIRequestPninetyfive{}\\
OpenAI & Incremental scoring & \DepthOpenAIScoringMean{} & \DepthOpenAIScoringMedian{} & \DepthOpenAIScoringPninetyfive{}\\
OpenAI & Cache lookup & \DepthOpenAICacheMean{} & \DepthOpenAICacheMedian{} & \DepthOpenAICachePninetyfive{}\\
OpenAI & Ranking assembly & \DepthOpenAIAssemblyMean{} & \DepthOpenAIAssemblyMedian{} & \DepthOpenAIAssemblyPninetyfive{}\\
OpenAI & Mixed query wall & \DepthOpenAIMixedMean{} & \DepthOpenAIMixedMedian{} & \DepthOpenAIMixedPninetyfive{}\\
OpenAI & Cache-only replay & \DepthOpenAIReplayMean{} & \DepthOpenAIReplayMedian{} & \DepthOpenAIReplayPninetyfive{}\\
\bottomrule\end{tabular}\end{table}

\FloatBarrier

\section{Discussion}\label{sec:discussion}
\subsection{Effectiveness within a shared retrieval protocol}
Both decision-based rerankers improve the observed dense baseline within the SciFact top-50 condition. This answers RQ2 at a fixed candidate stage. Richer query--document relevance scoring can improve ordering after representation-based retrieval. It does not show that either API independently discovers documents across the corpus. The shared \CandidateRecall{} candidate recall ceiling remains decisive because a reranker cannot recover missing evidence.

RQ1 isolates a different question. All official positives are guaranteed in the controlled pools. Sampled nonpositive documents are often unlike the harder dense-selected shortlist. Higher controlled metrics therefore cannot be interpreted as a gain from shrinking a production corpus, and no measured candidate-size scaling curve is established by placing the two conditions side by side. OpenAI's positive Recall point estimate relative to dense has an interval including zero, which narrows the effectiveness claim for that outcome.

Jev has higher observed Recall@5 and nDCG@10 than OpenAI in both tested conditions. The negative OpenAI-minus-Jev paired intervals support this descriptive ordering under the evaluated query sample, but the exploratory nature of the additional contrasts and unadjusted multiple comparisons limit confirmatory interpretation. The hybrid Recall interval's near-zero upper endpoint merits particular caution. Neither point estimates nor these intervals establish a universal best provider or superiority on other models, languages or application domains.

\subsection{Output saturation and interchangeable interfaces}
RQ3 shows that the numerical behavior of returned scores matters. OpenAI's many exact zeros produce large tied groups. A group entirely in a nonpositive tail may have no metric effect. Mixed-label groups crossing a cutoff can affect recall, and positional changes within a retrieved list can affect discounted gain. The saved tie realizations show measurable sensitivity, with primary results still defined by document-ID order. The realized ranges quantify this analysis only. They are not exhaustive bounds or alternative scores to select for reporting.

Jev also returns tied scores despite avoiding the exact endpoints in these runs. Endpoint fractions alone do not explain all ranking errors. Neither output distribution was calibrated against fully adjudicated relevance probabilities, and no architectural or training explanation follows from the histograms. The APIs implement semantically comparable relevance decisions, but replacing one with the other requires attention to model identity, input serialization, numerical behavior and request/response handling. A shared binary question is insufficient to make their probabilities interchangeable. Provider updates and nondeterministic serving may change outcomes even when model identifiers remain fixed.

\subsection{Practical implications and existing evidence}
Both APIs add hosted-request latency and monetary expenditure to local dense retrieval. Jev's observed condition costs and operational query medians are lower in these records, but dates, cache reuse, retries and timer boundaries differ. These observations do not estimate hardware-controlled inference efficiency. Overlapping cached pairs prevent adding usage-derived condition totals as fresh expenditure. Latency-tolerant ranking applications might accept this trade-off, whereas interactive applications would need deployment-specific validation. Neither application class was evaluated here.

Established cross-encoder and late-interaction work demonstrates query-dependent reranking \cite{nogueira2019bert,khattab2020colbert}. We compare Jev and OpenAI Decisions as independent relevance-scoring interfaces under the same candidate constraints, without inferring their internal architectures. No cross-encoder experimental baseline was included. Our contribution lies in a common SciFact protocol with retained query pairing, raw-score diagnostics and traceable costs/timers. The evidence answers the four questions within that protocol while leaving cross-domain robustness, calibration and alternative reranker comparisons unresolved.

\subsection{Candidate availability and the cost of deeper reranking}
RQ4 distinguishes the opportunity to retrieve known evidence from the scorer's ability to place it near the top. Extending the dense prefix recovers only \DepthAddedPairs{} additional known-positive pairs. Jev's final metrics rise slightly. OpenAI's Recall rises but nDCG falls slightly. All four depth intervals include zero, leaving small effects uncertain rather than demonstrating equivalence. Additional high-scored candidates can displace earlier documents, and incomplete judgments complicate interpretation of those intrusions. Coverage alone therefore does not determine final effectiveness.

The unique-condition API work and cost approximately double, while incremental Top-100 execution reuses historical scores. Mixed cached/new timers cannot quantify a cold latency penalty relative to Top-50. The observed quality--cost trade-off warrants caution about assuming deeper reranking is automatically beneficial. It does not establish Top-50 as universally optimal or Top-100 as useless because only two depths, one benchmark and one frozen dense retriever were evaluated. Controlled-100 is positive-enriched ranking. Its metrics do not measure the effect of expanding a dense shortlist.

\section{Conclusion}\label{sec:conclusion}
Under RQ1's controlled SciFact protocol, Jev and OpenAI Decisions both yield higher nDCG@10 point estimates than dense-only BGE-M3, and Jev has the highest observed Recall@5 and nDCG@10. The controlled OpenAI-minus-BGE Recall@5 interval includes zero, so that metric does not support an unqualified improvement claim.

For RQ2, both APIs improve ordering within the identical full-corpus BGE-M3 top-50 lists. nDCG@10 rises from \FullBGENDCG{} to \HybridOpenAINDCG{} with OpenAI and \HybridNDCG{} with Jev. Jev again has the highest observed metrics. The principal and exploratory unadjusted paired intervals remain scoped to this query sample. Neither API replaces the first-stage dense retriever or recovers documents absent from its candidates.

For RQ3, OpenAI's returned score saturation and exact ties introduce measurable ranking sensitivity. Primary deterministic results remain unchanged, and the secondary seeded ranges do not constitute exhaustive bounds. Both hosted scorers add operational latency and usage-derived expenditure under different execution conditions. This single-dataset evidence does not identify a universal best decision model. Direct probabilistic relevance decisions can complement dense retrieval, but scorer selection should account for ranking effectiveness, score behavior, operational latency and cost.

For RQ4, increasing the saved BGE-M3 prefix from 50 to 100 improves known-positive candidate recall but changes final Recall@5 and nDCG@10 only modestly. All four exploratory paired depth intervals include zero, so a positive depth effect is not clearly established and equivalence is not demonstrated. Jev retains the highest observed full-corpus metrics at Top-100. The cross-provider Recall interval includes zero. Unique-condition relevance work and API cost approximately double, with incremental expenditure reduced by compatible cache reuse. These findings support evaluating candidate coverage, scorer behavior and operational cost jointly within a specific protocol.

\FloatBarrier
\begingroup\raggedright
\endgroup

\end{document}